\documentclass[aps,pre,twocolumn,showpacs,floatfix]{revtex4}
\usepackage[dvips]{graphicx}
\usepackage{bm,amsmath}

\begin{document}
\title{A study  of polymer knots using a
simple knot invariant written consisting of multiple contour integrals}
\author{Yani Zhao}
\email{yanizhao@fermi.fiz.univ.szczecin.pl}
\author{Franco Ferrari}
\email{ferrari@fermi.fiz.univ.szczecin.pl}
\affiliation{CASA* and Institute of Physics, University of Szczecin,
  Szczecin, Poland} 
\date{\today}

\begin{abstract}
In this work the thermodynamic properties of short polymer knots (up
to 120 segments) defined on a simple cubic lattice are studied with
the help of the Wang-Landau Monte Carlo algorithm. The sampling
process is performed using pivot transformations starting from a given
seed conformation. Both cases of short-range attractive
and repulsive interactions acting on the monomers are considered. The
properties of the specific energy, heat capacity and gyration radius
of the knots $3_1,4_1$ and $5_1$ are discussed. It is found that the
heat capacity exhibits a sharp peak. If the interactions
are attractive, similar peaks have been observed also in single open
chains and have been related to the
transition from a frozen crystallite state to an expanded coil
state. 
Some other peculiarities of the behavior of the analyzed observables
are presented, like for instance the increasing or
decreasing of the knot specific energy at high temperatures with
increasing polymer lengths depending if the interactions are
attractive or repulsive. 
Besides the investigation of the thermodynamics of 
polymer knots,  the second goal of this paper is to introduce a method
for distinguishing the topology of a knot based on a topological
invariant
which
is in the form of multiple contour integrals and explicitly depends on
the physical trajectory of the knot. The chosen
invariant, denoted here $\varrho(C)$, is related to the second
coefficient of the Conway polynomial. It has been  first isolated from
the amplitudes of a Chern-Simons field theory with gauge group
$SU(N)$. It is shown that this invariant is very reliable in
distinguishing the topology of polymer knots.   
One of the advantages of the proposed approach is that it allows to
reduce the number of samples needed by the Wang-Landau algorithm. 
Some solutions to speed up the calculations 
of $\varrho(C)$ 
exploiting Monte Carlo integration techniques are developed.
\end{abstract}
\maketitle
\section{Introduction}\label{introd}
Long polymers are very likely to be found in the configuration of
knots or links. The topological properties of polymers with closed
conformations play indeed an 
important role in physics, chemistry and biology. For that reason,
they are being actively investigated \cite{grosberg,dna,katritch96,katritch97,krasnow,
laurie,cieplak,marko,
liu,wasserman,sumners,vologodski,orlandini,
marenduzzoorlandinimichelettiphysrep, levene,
kurt,mehran,pp2,yan,arsuaga,arsuaga2,
metzler,pieranski,diao,sumners2,faddeev}. A particularly
challenging problem is that of the statistical mechanics of polymer
knots. Up to now, a satisfactory analytical model exists only in the
case of two polymer rings linked together \cite{FFIL,FFIL2}, but there
is no analogous 
model for a knot despite many attempts, see for instance
\cite{kholodenkovilgis,FFAP,FFNOVA,kleinert} for a review on  this
subject. Moreover, the scaling laws of  
the most important 
observables of polymer knots, like for instance the gyration radius,
are still a subject of intense research
\cite{grosbergjpa,stella,quake,mansfielddouglas,suzukietal}.

The main difficulty behind the treatment of polymer knots, as well as
polymer links, is to distinguish the wealth of different topological
configurations of such systems. This problem arises
in analytical models because it is necessary  to impose
topological constraints in 
order to avoid that the statistical fluctuations affect the initial
topological state. This is a physical requirement, dictated by the
fact that, once a 
polymer knot or link has been formed, its topological state cannot be
modified
 without
breaking the covalent bonds holding together the monomers.
Thus, unwanted changes of the topological configuration must be
detected and rejected. 
In numerical simulations, instead, a widely used way to generate
polymer knots or links is to consider self
avoiding walks (SAW's) that, at a certain point, intersect themselves
forming closed conformations \cite{vologodski,descloizeaux,hazi,michels}. The problem in this case
 is to sort out the
topological configurations that are of interest
from all the other configurations produced within this approach.
This process of generating knot and links is not very efficient for
our purposes,
because the probability of formation of knots or links of a given type
from SAW's is
very low, see for instance
Ref.~\cite{probability1,probability2,probability3}. 
Alternatively, it is possible to start from a polymer ring with a seed
trajectory that is out of equilibrium, but already in the desired
topological configuration. Later, the system is
equilibrated using the 
so-called pivot transformations \cite{pivot,Rensburg}. There exist
pivot transformations 
that are automatically preserving the topological state of the
system,
see for example \cite{swetnam}, however they are
able to modify only a very small part of the whole polymer, a fact
that considerably increases the time for reaching the equilibrium,
especially for very long polymers. 
In \cite{yzff} another method
has been proposed, called the Pivot Algorithm and Excluded Area (PAEA)
method, in which more general pivot 
transformations are considered, including those that
could potentially change the topology. In this case, large
transformations are not easy to be implemented.

In conclusion, apart from a few exceptions, like for instance the
already cited works of 
\cite{swetnam,yzff} and notably the dynamic Monte Carlo approach
\cite{dmca}, 
that however leads to some level of polydispersity,
in most numerical computations involving polymer
knots, topological invariants are exploited
whenever it becomes necessary to distinguish the topological
configuration of the system under investigation.
The most
popular topological invariants are given in the form of polynomials or
of multiple contour integrals computed along the physical trajectories
of the polymers. The latter invariants are easier to be used in
analytical models than the former, because the
coefficients of the polynomials are not directly related to
the polymer conformation. The particular simplicity of the Gauss
linking number, which consists in a double contour integral, is the
main reason for which it has been possible to derive an analytical
model of two linked polymer rings \cite{FFIL,FFIL2,FFIL3}. 
Moreover, the Gauss linking number has already
been used in numerical simulations of polymer systems, see for
instance \cite{ralf,Kremer1}.
Unfortunately, in order to distinguish the topology of knots,
there is no simple invariant like the Gauss linking number. So far,
the statistical properties of polymer knots have been studied
numerically with
the help of topological invariants like the Alexander polynomials
\cite{vologodski} or 
the HOMFLY polynomials \cite{michalke}, see the detailed
textbook by Kleinert \cite{kleinert} for an extensive review
on that subject. Of course, there is
a plenty of other  knot 
invariants that have been or could be applied, for instance the Conway
polynomials \cite{conway}, the Arf-Casson invariant
\cite{GMMknotinvariant} or the Milnor \cite{milnor} and 
 Vassiliev-Kontsevich
invariants \cite{kontsevich}. The latter three have an explicit
representation in terms of 
multiple integrals computed along the knot trajectories.

The purpose of this paper is to show that
knot invariants that are given in the form of  multiple contour
integrals can represent a valid alternative in numerical calculations  
to polynomial knot invariants or even to the PAEA method,
which is able to detect the topology changes exactly
 and
has been proven to allow very fast computations of the statistical
properties of polymer knots \cite{yzff,yf}.
In particular, we will concentrate on a topological invariant
denoted here $\varrho(C)$, where $C$ denotes the trajectory
of the knot. $\varrho(C)$  has been derived from the one-loop
amplitudes
of non-abelian Chern-Simons field theories with gauge group $SU(N)$
\cite{GMMknotinvariant}.
It is related to the Arf-Casson invariant and to the second
coefficient of the Conway polynomials \cite{GMMknotinvariant}. Its value 
can be analytically computed for any given knot configuration.
$\varrho(C)$ is the simplest knot invariant represented in terms of
multiple contour integrals.

The knot invariant $\varrho(C)$ 
is applied here in order to derive the average values of the
specific energy, heat capacity and gyration radius 
of several different knot configurations by means of the Wang-Landau
algorithm \cite{wanglandau}.  
We find in this way that the examined knots, corresponding namely to
the trefoil $3_1$, the figure-eight $4_1$ and
$5_1$\footnote{We use here the Alexander-Briggs notation of knots,
  see for example \cite{kleinert} for its explanation.}, undergo with the
temperature a phase transition, which is probably
from a frozen crystallite state to an expanded coil state
similarly to what
happens in the case of a single polymer chain proposed
in~\cite{binder}. Other physical properties of polymer knots are discussed.
Compared with the PAEA method, 
the use of $\varrho(C)$
allows to reduce
the number of samples necessary for the calculations of the averages
of the 
observables with the Wang-Landau algorithm. 
As it will be seen, this reduction 
is due to the fact that with $\varrho(C)$, large pivot transformations
can be exploited which are able to change relevant portions of the
knot. In this way, the exploration of the whole set of available
conformations becomes faster.
Despite the decreasing of the number of samples,
the computations last in general
longer than those performed with the PAEA method, because   the
expression of $\varrho(C)$ contains
 quadruple integrals that should be evaluated numerically and this
 takes some time. As a consequence, we present here the results for
 relatively short polymer knots up to
 $L=120$, where $L$ is the number of segments composing the knot
on a simple cubic lattice. Actually, there is no problem in studying
longer polymer knots. The reason is that invariants given in the form
of contour integrals can be computed for arbitrarily deformed knots,
not necessarily defined on a lattice, provided the topological
configuration remains the same after the deformation. Thanks to this
fact, we have found that it is possible to shorten the number of
segments by a factor 
three, speeding up the calculation of $\varrho(C)$ considerably.
Even using this trick, the method requires
 too long times (exceeding a month on a modern workstation) if $L>360$.
The use of $\varrho(C)$ becomes however competitive in the
equilibration of very long polymers because, as already mentioned, it offers
the possibility of exploring very fast the set of 
conformations compatible with the topological constraints.

The rest of the current work is organized as follows. Our simulation
set-up and the topological
invariant $\varrho(C)$ are introduced in Section~\ref{SectionII}. The Monte Carlo
integration method we use to calculate the knot invariant $\varrho(C)$ is introduced in
Section~\ref{sectionIII}. An explanation of the Wang-Landau algorithm
with particular attention to its applications to the statistical
mechanics of polymers is provided in Section~\ref{sectionIV}. The results on the thermal properties of
different polymer knots are presented in Section~\ref{sectionV}. We also
compare our calculations with the results obtained with the PAEA method
in Refs.~\cite{yzff,yf}. Finally, in
Section~\ref{sectionVI} 
we draw our conclusions and possible generalizations of this work are
briefly discussed.

\section{Methodology}\label{SectionII}
\subsection{General outline of the used methodology}
First of all, a brief digression on the used terminology is in order.
Throughout this work the word configuration refers to a
particular topological state of a polymer knot.
The word conformation will instead denote the particular shape in the
space of the trajectory of a polymer knot in a given topological configuration.

At this point it is possible to go back to the outline of the
methodology.
We adopt the strategy of considering at the beginning a
seed conformation of the knot to be analyzed. The knot is a
self-avoiding polygon 
defined on a simple cubic lattice with edges of unit length. The
monomers are located on the vertices of the lattice. 
Let $L$ denote the total length of the polygon. Since it consists of
edges or segments of unit length, $L$ coincides also with the number
of segments composing the polygon.
The crossing of the knot trajectory with itself at some point on the
lattice is forbidden.
The starting seed configuration is equilibrated and then used to
compute the thermodynamic properties of the studied knot.
The relevant observables are calculated by means of the Wang-Landau
Monte Carlo algorithm \cite{wanglandau}, which will be explained in
some more details later. Both cases of attractive and repulsive
interactions are considered.
For the equilibration of the knot and  the
sampling in the Wang-Landau algorithm, the polymer conformations are 
randomly modified by exploiting the pivot transformations described in
Ref.~\cite{pivot}. 
These  transformations can involve any number $N$
of segments such that $1<N\le L$ and are not preventing the
crossing of the lines of the knot trajectory $C$.
For that reason, the invariant 
$\varrho(C)$ should be applied in order to check if the topology of
the knot has been 
altered after each pivot transformation. If this is the case, the
transformation is rejected and a new one is considered.
\subsection{The topological invariant $\varrho(C)$}
To avoid topology changes
of a polymer knot $C$ potentially occurring after the pivot
transformations, we 
will use in this work
a topological invariant
that has been
derived from the one-loop amplitude of the Wilson loop in 
non-abelian $SU(N)$ Chern-Simons field theories
\cite{GMMknotinvariant,LR,cotta}. 
The most important characteristic of this invariant, that will be
denoted $\varrho(C)$, is that it can be expressed in the form of a
sum of multiple contour integrals:
\begin{equation}
\varrho(C)=\varrho_{1}(C)+\varrho_{2}(C) \label{z1}
\end{equation}
where the knot $C$ is represented as an oriented closed path of length
$L$. The contribution $\varrho_1(C)$ is given by the triple integral:
\begin{equation}
\varrho_{1}(C)=-\frac{1}{32\pi^3}\oint_{C}dx^{\mu}\int^{x}dy^{\nu}\int^{y}
dz^{\rho}I_{\mu\nu\rho}(\vec{x},\vec{y},\vec{z}), \label{z2}
\end{equation}
with
\begin{eqnarray}
I_{\mu\nu\rho}(\vec{x},\vec{y},\vec{z})&&=\epsilon^{\alpha\beta\gamma} 
\epsilon_{\mu\alpha\sigma}  
\epsilon_{\nu\beta\lambda}\epsilon_{\rho\gamma\tau}\nonumber\\&& \times\int d^{3}\vec{\omega} \frac{(\omega -x)^\sigma}{|\vec{\omega} -\vec{x}|^3} 
\frac{(\omega -y)^\lambda}{|\vec{\omega} -\vec{y}|^3} \frac{(\omega -z)^\tau}{|\vec{\omega} -\vec{z}|^3} \label{z3}
\end{eqnarray}
while the second part $\varrho_{2}(C)$ is:
\begin{eqnarray}
\varrho_{2}(C)&=&\frac{1}{8\pi^2}\oint_{C}dx^{\mu}\int^{x}dy^{\nu}\int^{y}dz^{\rho}
\int^{z}dw^{\sigma}\epsilon_{\sigma\nu\alpha}\epsilon_{\rho\mu\beta}
\nonumber\\&& \times
\frac{(w-y)^\alpha}{|\vec{w}-\vec{y}|^3} \frac{(z-x)^\beta}{|\vec{z}-\vec{x}|^3} \label{z4}
\end{eqnarray}
In the above formulas greek letters denote space indexes.
The variables $x^\mu, y^\nu, z^\rho$ and $w^\sigma$,
$\mu,\nu,\rho,\sigma=1,2,3$, are the components of the
radius vectors describing the positions of four points on the same
curve $C$. 
$\epsilon_{\mu\nu\rho}$ represents instead the completely
antisymmetric tensor uniquely defined by the condition
$\epsilon_{123}=1$. 
 The integrations along the path $C$ in (\ref{z2}) and
(\ref{z4}) are path ordered. This can be seen explicitly by
 parametrizing the trajectory $C$ with the arc-length:
\begin{widetext}
\begin{eqnarray}
\varrho_{1}(C)&=&-\frac{1}{32\pi^3}\int_0^L ds\frac{dx^{\mu}(s)}{ds}\int_0^sdt
\frac{dy^{\nu}(t)}{dt}\int_0^tdu\frac{dz^{\rho}(u)}{du}
I_{\mu\nu\rho}(\vec{x},\vec{y},\vec{z}), \label{z2po}
\end{eqnarray}
and
\begin{eqnarray}
\varrho_{2}(C)&=&\frac{1}{8\pi^2}
\int_0^L ds\frac{dx^{\mu}(s)}{ds}\int_0^sdt
\frac{dy^{\nu}(t)}{dt}\int_0^tdu\frac{dz^{\rho}(u)}{du}\int_0^udv
\frac{dw^{\sigma}(v)}{dv}
\epsilon_{\sigma\nu\alpha}\epsilon_{\rho\mu\beta}
\frac{(w(v)-y(t))^\alpha}{|\vec{w}(v)-\vec{y}(t)|^3} \frac{(z(u)-x(s))^\beta}{|\vec{z}(u)-\vec{x}(s)|^3} \label{z4po}
\end{eqnarray}
\end{widetext}
It has been shown that the knot invariant appearing above is related
to the second 
coefficient $a_2(C)$ of the Conway polynomial of a knot $C$ through
the following 
relation \cite{GMMknotinvariant}: 
\begin{equation}
a_2(C)=\dfrac{1}{2}\left[\varrho(C)+\dfrac{1}{12}\right]\label{rela2rho}
\end{equation}
The Conway polynomials are well known and their coefficients can be
computed analytically for every knot topology, so that thanks to
Eq.~(\ref{rela2rho}) it is easy to derive also the values
of $\varrho(C)$. In Table~\ref{table:conway} we give a list of the
second coefficients of the Conway 
polynomial and the corresponding values of $\varrho(C)$  for
the knot configurations that will be studied here.

\begin{table}[ht]
\caption{ This table provides the values of the second coefficients of
  the Conway polynomials and of
the corresponding topological
  invariants for the trefoil $3_1$, the figure-eight
  $4_1$ and the knot $5_1$.} 
\centering
\begin{tabular}{c c r}
\hline\hline
 knot type & $a_2(C)$& $\varrho(C)$\\ [0.5ex]
\hline
$3_1$ &1   &$+\frac{23}{12}$   \\
$4_1$ &-1  &$-\frac{25}{12}$  \\ 
$5_1$ &3   &$+\frac{71}{12}$   \\ [1ex]
\hline
\end{tabular}
\label{table:conway}
\end{table}
$\varrho(C)$ is the simplest  known knot invariant that can
be expressed in the form of contour integrals.
Like any other knot invariant,
$\varrho(C)$ is not able to distinguish different knots unambiguously. 
For example,
the trefoil knot $3_1$ has $\varrho(C)=\frac{23}{12}$, exactly the
same value of the 
knots $6_3,7_6,8_{13}$ and many others.
  However, we should keep in mind
that the main role of a knot invariant in studying the
thermal and mechanical properties of polymer knots is not to guess its
topological configuration.
In fact, the topological configuration is
known since the beginning. The problem is rather to preserve that
configuration against thermal fluctuations, because without any constraint
the polymer trajectories are allowed to cross themselves, a fact that
can potentially alter a knot. 
 The probability that due to thermal fluctuations
a polymer ring jumps from one knot configuration to another with the
same value of  $\varrho(C)$ is very low, as it has been observed in our
simulations. 
What emerges from them is that it is very unlikely that a knot
passes to another configuration with the same value of 
$\varrho(C)$ after a pivot transformation. 
Most probably, one ends up with the 
trivial knot or, somewhat less
frequently, with a conformation with lower number of crossings $C'$
such that $\varrho(C)\ne\varrho(C')$.
For the purposes of this work, it is thus possible to affirm that
$\varrho(C)$ is a
powerful knot invariant. To convince oneself, it is sufficient to recall that,
if one considers the simplest knots up to ten crossings, there are
particular topological configurations that are uniquely distinguished
by $\varrho(C)$, like $9_1$
and $10_3$ or are very efficiently distinguished from all the others
because their 
corresponding values of  $\varrho(C)$ occur rarely.
Luckily, if the values of $\varrho(C)$ for two topologically different
knots are not the same, the
smallest difference between  them
is $2$.
As an example, the knot  $5_1$ has
$\varrho(C)=\frac{71}{12}$, while for the topologically inequivalent knots
$5_2$ and $9_{11}$ we have $\varrho(C)=\frac{47}{12}$ and
$\varrho(C)=\frac{95}{12}$ respectively. 
This allows to choose the number of sampling points in such a way that
the variance
in the Monte
Carlo evaluation  of $\varrho(C)$ 
is low enough that the probability of confusing  
two different knot topologies due to numerical errors is negligible.

The price to be paid for this
efficiency in distinguishing knots is the complicated expression of
$\varrho(C)$. The most time-consuming contribution to $\varrho(C)$
is the quadruple contour integral necessary to compute
$\varrho_2(C)$ in Eq.~(\ref{z4}). For a knot of length $L$, the
evaluation
time of  $\varrho(C)$ scales as $L^4$.
This is approximately 
one order more than the time necessary to evaluate the 
Alexander polynomial of a knot \cite{vologodski,deguchitsurusaki,Muthukumar},
which scales as $(M-1)^3$. Here $M$ denotes the number of crossings
which is necessary to represent the knot  by projecting it on an
arbitrary plane, see \cite{Muthukumar} for more details.
Of course, also the computation of the Alexander polynomial
becomes prohibitive for polymers which are long or have compact
conformations, because in these cases the number of crossings $M$
drastically increases
\cite{marenduzzoorlandinimichelettiphysrep}. Moreover, 
the scaling law $(M-1)^3$ of the computational time is true only if
the determinant
of a $M\times M$ matrix that arises in the algorithm for computing the
Alexander polynomial is evaluated with the method of Gaussian elimination, which
is subjected on round-off error that become important when $M$ is large.

One advantage of $\varrho(C)$ is that its calculation can be extended
without any effort to any
kind of trajectory, not necessarily on a cubic lattice. This fact is very
helpful when the polymer is long, so that it is advisable to
decrease the number  $L$ of its segments.
In a very simple way it is possible to reduce $L$ 
by
a factor three
without destroying
the topology  by replacing in the knot
every group of three contiguous segments with a single segment. 
As well, there is no problem in distinguishing the changes of topology
when the number $N$ involved in the pivot transformations becomes
large.

Like the
Alexander polynomial, also 
$\varrho(C)$ is not able to distinguish uniquely  two different
topological configurations and
is subjected to numerical errors. However, we have seen above that, for the
goals of this work, the invariant $\varrho(C)$ is powerful enough.

\section{Monte Carlo evaluation of path ordered contour integrals on a
lattice} \label{sectionIII}
\subsection{Simpson's rule vs. Monte Carlo method}
\begin{figure}
\begin{center}
\includegraphics[width=3in]{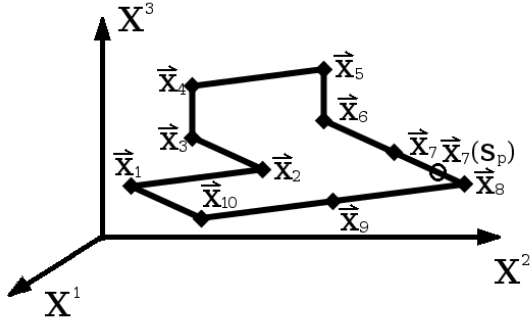}
\caption{\label{conf}
This figure illustrates the notation used in
Section~\ref{sectionIII} to describe closed
contours on a simple cubic lattice with the help of the example of a 
short contour of length $L=10$. The coordinate of the
point $p$, represented in the figure by an empty circle, is 
$\vec{x}_7(\tilde s_p)=(x^1_7(\tilde s_p),x^2_7(\tilde s_p),
x^3_7(\tilde s_p))$, where 
 $\tilde s_p$ is the distance of the point $p$
from the lattice site $\vec{x}_7$.}  
\end{center}
\end{figure}

First of all, we
introduce some notation that will be useful in this Section.
The contour $C$ describing the physical trajectory of the knot in
space is represented here as a curve $\vec{x}(s)$, with $0\le s\le L$.
Of course,
$C$ is consisting of a set of discrete segments, see
Fig.~\ref{conf} for an example with $L=10$, and this fact 
should be taken into account.
To this purpose, let's denote
with
$\vec{x}_i$, $i=1,\cdots, L$, the locations of the lattice sites
through which the closed contour $C$ is passing. The $i$-th segment of
the loop $C$ forms a vector $\vec{x}_{i+1}-\vec{x}_i$ for
$i=1,\cdots,L-1$. The $L$-th segment is instead  associated with the
vector
$\vec{x}_{1}-\vec{x}_L$. 
Next, let $\vec{x}_i (\tilde s)$, with $i=1,\cdots,L$, be the restriction of
the curve $\vec {x}(s)$ to the $i-$th segment. Here we have introduced
the segment's arc-length $\tilde s$ such that $0\le \tilde s\le 1$.
On the $i-$th segment, $\tilde s$ is related to $s$ 
by the formula
\begin{eqnarray}
\tilde s = s-i\qquad&\mbox{for}&\qquad i=1,\ldots,L-1\\
\tilde s= s\qquad&\mbox{for}& i=L
\end{eqnarray}
Explicitly, the expression of $x_i(\tilde s)$ is given by:
\begin{eqnarray}
\vec{x}_i(s)&=&\vec{x}_i+\tilde s(\vec{x}_{i+1}-\vec{x}_i)
\qquad i=1,\ldots,L-1\label{disccont1}\\
\vec{x}_L(s)&=&\vec{x}_L+\tilde s(\vec{x}_{1}-\vec{x}_L)
\label{disccontL}
\end{eqnarray}
From the above equations, it is easy to derive also the derivative
of $\vec{x}_i(\tilde s)$ with respect to $\tilde s$.
Whenever it will be necessary to specify points on different elements of
the loop $C$, for instance on 
segments $i,j,k,l,\ldots$, they will be denoted with the symbols
$\vec{x}_i(\tilde s), \vec{y}_j(\tilde t), \vec{z}_k(\tilde u),
\vec{w}_l( \tilde v),\ldots$,
where $i, j, k, l=1,\cdots,L$ and $0\le s, t, u, v\le 1$. 

At this point we are ready to rewrite the quantities $\varrho_1(C)$ and
$\varrho_2(C)$ displayed in Eqs.~(\ref{z2po}) and (\ref{z4po})
respectively in a form that is  suitable for applying
the standard Simpson's rule:
\begin{widetext}
\begin{eqnarray}
\varrho_{1}(C)&=&-\frac{1}{32\pi^3}\sum_{i=1}^L\sum_{j=1}^i\sum_{k=1}^j\int_0^1
d\tilde s \frac{dx^{\mu}_i(\tilde s)}{d\tilde
  s}\int_0^{1-\delta_{ij}(1-\tilde s)}
d\tilde t \frac{dy^{\nu}_j(\tilde t)}{d\tilde t}\nonumber\\
&\times&\int_0^{1-\delta_{jk}(1-\tilde t)}d\tilde u
\frac{dz^{\rho}_k(u)}{d\tilde u} 
I_{\mu\nu\rho}(\vec{x}_i(\tilde s),\vec{y}_j(\tilde t),\vec{z}_k(
\tilde u)) \label{rhooneext}
\end{eqnarray}
and
\begin{eqnarray}
\varrho_{2}(C)&=&\frac{1}{8\pi^2}
\sum_{i=1}^L\sum_{j=1}^i\sum_{k=1}^j\sum_{l=1}^k\int_0^1 d\tilde s
\frac{dx^{\mu}_i(\tilde s)}{d\tilde s}\int_0^{1-\delta_{ij}(1-\tilde s)}
d\tilde t
\frac{dy^{\nu}_j(\tilde t)}{d\tilde t}
\int_0^{1-\delta_{jk}(1-\tilde t)}d\tilde u\frac{dz^{\rho}_k(\tilde
  u)}{d\tilde u}
\nonumber\\& \times&
\int_0^{1-\delta_{kl}(1-\tilde u)}
d\tilde v\frac{dw^{\sigma}_l(\tilde v)}{d\tilde v}
\epsilon_{\sigma\nu\alpha}\epsilon_{\rho\mu\beta}
\frac{(w_l(\tilde v)-y_j(\tilde t))^\alpha}
{|\vec{w}_l(\tilde v)-\vec{y}_j(\tilde t)|^3}
\frac{(z_k(\tilde u)-x_i( \tilde s))^\beta}{|\vec{z}_k(\tilde
  u)-\vec{x}_i( \tilde s)|^3}  \label{rhotwoext} 
\end{eqnarray}
\end{widetext}
The boundaries in the integrals of $\varrho_1(C)$ and $\varrho_2(C)$
have been chosen in such a way that the path ordering of the
trajectories is preserved. For example, if the integrals over $\tilde s$ and
$\tilde t$ are performed over two different segments, then the boundaries for
both variables are ranging between $0$ and $1$. Instead, if the
integrals are performed on the same segment $i=j$, then it must be
that $0\le \tilde s\le 1$ and $0\le \tilde t\le \tilde s$.
Let us note that the integrals over the variable $\tilde v$ in
Eq.~(\ref{rhotwoext}) can be 
performed exactly. 
The  remaining integrals over
$\tilde s,\tilde t$ and $\tilde u$ should be evaluated numerically,
for instance by 
means of the Simpson's rule.
It turns out that for the evaluation of these
integrals, the zeroth order approximation, obtained by
substitutions of the kind: 
\begin{eqnarray}
&&\int_0^{1-\delta_{ij}(1-\tilde s)}dy_j^{\nu}(\tilde
  t)f_{\nu}(\vec{y}_j( \tilde t))\nonumber\\&&
 \sim (y_j^{\nu}(1)-y_j^{\nu}(0))
\frac{f_{\nu}(\vec{y}_j(1))+f_{\nu}(\vec{y}_j(0))}{2} \label{crudeapprox}
\end{eqnarray}
is sufficient and gives satisfactory results. The problem is that,
even exploiting 
the 
crude approximations of Eq.~(\ref{crudeapprox}) in
order to distinguish the topology changes of a given polymer knot,
still a sum of $\frac{L^4}{24}$ terms should be evaluated in order to derive
the value of $\varrho_2(C)$ from Eq.~(\ref{rhotwoext}). Already in the
case of polymers of length $L=100$ or more, this number becomes prohibitively
high for practical purposes. In fact,  
the  
Wang-Landau procedure used to compute the density of states requires
several millions of samples to be evaluated. For that reason,
it is much better to estimate $\varrho_1(C)$ and $\varrho_2(C)$
performing the
integration with Monte Carlo techniques. 

The idea is to regard the contour integrals in Eqs.~(\ref{z2po}) and
(\ref{z4po}) as usual multiple integrals over the variables $s, t, u$
and $v$: 
\begin{equation}
\varrho_1(C)=\int_0^Lds \int_0^sdt\int_0^t du F_1(s,t,u) \label{rhoonemi}
\end{equation}
and
\begin{equation}
\varrho_2(C)=\int_0^Lds \int_0^sdt\int_0^t du\int_0^u dv F_2(s,t,u,v) 
\label{rhotwomi}
\end{equation}
where
\begin{eqnarray}
F_1(s,t,u)&=&-\frac{1}{32\pi^3}\frac{dx^{\mu}(s)}{ds}\frac{dy^{\nu}(t)}{dt}
\frac{dz^{\rho}(u)}{du}\nonumber\\&&
\times I_{\mu,\nu,\rho}(\vec{x}(s),\vec{y}(t),\vec{z}(u))
\end{eqnarray}
and
\begin{eqnarray}
&&F_2(s,t,u,v)=\frac{1}{8\pi^2}\frac{dx^{\mu}(s)}{ds}\frac{dy^{\nu}(t)}{dt}\frac{dz^{\rho}(u)}{du}\frac{dw^{\sigma}(v)}{dv}\nonumber\\&&
\times\epsilon_{\sigma\nu\alpha}\epsilon_{\rho\mu\beta} \frac{(w(v)-y(t))^\alpha}
{|\vec{w}(v)-\vec{y}(t)|^3}
\frac{(z(u)-x(s))^\beta}{|\vec{z}(u)-\vec{x}(s)|^3} 
\end{eqnarray}

The variables $s, t$ and $u$ in Eq.~(\ref{rhoonemi}) span a space of
volume $V_1=\frac{L^3}{6}$, while the variables $s, t, u$ and $v$ in
Eq.~(\ref{rhotwomi}) span a space of volume $V_2=\frac{L^4}{24}$. To
evaluate the right hand sides of Eqs.~(\ref{rhoonemi}) and
(\ref{rhotwomi}) via Monte Carlo integration, we can exploit the
general formula  
\begin{widetext}
\begin{eqnarray}
\int_{a_1}^{b_1}d\xi_1\int_{a_2}^{\xi_1}d\xi_2\cdots\int_{a_m}^{\xi_{m-1}}
d\xi_m f(\xi_1,\cdots ,\xi_m) 
\approx \dfrac{1}{N}\left[\sum_{i=1}^N f(\xi_1^{(i)},\cdots ,\xi_m^{(i)})(b_1-a_1)\prod_{\sigma=2}^m (\xi_\sigma^{(i)}-a_\sigma)\right]
\end{eqnarray}
\end{widetext}
where the $\xi_{\sigma}^{(i)}$'s, $i=1, \cdots, N$ and
$\sigma=1,\cdots,m$ denote randomly chosen variables in the range: 
\begin{equation}
\begin{array}{ccl}
[a_1,b_1]&\mbox{when}&\sigma=1\\
{[a_{\sigma},\xi_\sigma]}&\mbox{when}&\sigma=2,\ldots,m
\end{array}
\end{equation}

%
In the numerical evaluation of $\varrho(C)$, 
we consider the trajectory $C$ of the knot, which in principle is a
polygon on a simple cubic lattice, as a continuous curve
$\vec{x}(s)$.
For a given value of $s$, the segment on which
the point $\vec{x}(s)$ is located is identified by the relation
\begin{eqnarray}
i&=&[s]+1 
\end{eqnarray}
where $[s]$ denotes the integer part of  $s$.
The components of the curve $\vec{x}(s)$ 
are obtained using its
restriction $\vec{x}_i(\tilde s)$ to the $i-$th segment,
whose components can be computed
exploiting Eqs.~(\ref{disccont1}) and (\ref{disccontL}).
The components of $\vec{y}(t)$,  $\vec{z}(u)$ and  $\vec{w}(v)$
are derived analogously.

We stress the fact that the integrands $F_1(s,t,u)$ and $F_2(s,t,u,v)$
are regular, even if for instance $F_2(s,t,u,v)$ seems
to be divergent when $\vec{w}(v)=\vec{y}(t)$ or
$\vec{z}(u)=\vec{x}(s)$. 
Analytically, it is possible to prove that these singularities cancel as
expected in a topological invariant.
In numerical
computations the situation looks however different, because one has to
cope with terms that are separately diverging, but whose sum is finite.
%
%
 A regularization is thus necessary in order to eliminate these
ambiguities. 
Let us first consider
 the computation of $\varrho_2(C)$. Here there are potential
problems whenever 
\begin{eqnarray}
\vec{w}(v)-\vec{y}(t)=0 &\mbox{or}& \vec{z}(u)-\vec{x}(s)=0. \label{rhotwosing}
\end{eqnarray}
However, the probability of the occurrence of such situations by choosing
randomly the variable $s,t,u,v$ is very low. In fact, if the
calculations are performed using
double precision variables, the number of digits after the floating
point is so high, that in 
practice the conditions displayed in Eq.~(\ref{rhotwosing}) are never
satisfied. More serious is the case of $\varrho_1(C)$. After the
integration over $\vec \omega$ in Eq.~(\ref{z3}), we get the explicit
expression of $F_1(s,t,u)$ which is reported in Appendix A. We see that,
besides the singularities at the points satisfying the conditions:
\begin{eqnarray}
\vec{y}(t)=\vec{x}(s)\qquad \vec{z}(u)=\vec{x}(s)\qquad
\vec{y}(t)=\vec{z}(u)\label{rhoonetriv} 
\end{eqnarray}
it appears also a pole whenever the equation
\begin{equation}
A=|\vec{y}(t)-\vec{x}(s)||\vec{z}(u)-\vec{x}(s)|+(\vec{y}(t)-\vec{x}(s))\cdot(\vec{z}(u)-\vec{x}(s))=0 \label{rhoonenontriv}
\end{equation}
is verified.
While it is very unlikely  that, during the random sampling,
the divergences of Eq.~(\ref{rhoonetriv}) will appear
due to the same reasons explained in the case of the analogous
divergences in Eq.~(\ref{rhotwosing}), the condition
(\ref{rhoonenontriv}) is very easy to be realized. It is sufficient
for example that the two vectors 
$\vec{y}(t)-\vec{x}(s)$ and $\vec{z}(u)-\vec{x}(s)$ have opposite
orientations and two zero components. Depending on the topology of the
knot, this situation may occur rather often. Even if the number of
cases in which this happens is negligible with respect to the total number
of sampled points generated during the Monte
Carlo computation of 
$\varrho_1(C)$, still the result may be spoiled by overflow or
underflow errors. To cure the singularity in Eq.~(\ref{rhoonenontriv})
when the quantity $A$ is equal to zero, 
we use the framing regularization introduced in
\cite{Witten}.
This consists in performing an almost infinitesimal
shift of the curves $\vec{x}(s)$, $\vec{y}(t)$,
$\vec{z}(u)$ and $\vec{w}(v)$
along a direction which is normal 
to the knot $C$. By denoting with $\vec{n}(s)$ the unit vector that gives the
normal direction to $C$, this means to replace for instance
$\vec{x}(s)$ with the
quantity
$\vec{x}(s)+\epsilon \vec{n}(s)$, where $\epsilon$ is very
small, let's say of the order $\epsilon\sim 10^{-10}$. This is
sufficient to eliminate 
all singularities occurring when the condition
(\ref{rhoonenontriv}) is fulfilled while preserving the topological
properties of $\varrho(C)$ as shown in Ref.~\cite{GMMknotinvariant}.
We have checked that the result of the
calculation of $\varrho_1(C)$ is not much sensitive to the value of
$\epsilon$. 

With  the above setup, we have found that a number of Monte
Carlo samples of the order of a few millions is sufficient to evaluate
both contributions $\varrho_1(C)$ and $\varrho_2(C)$ to the knot
invariant $\varrho(C)$ with a satisfactory precision for polymers of
length
up to $L=125$.
Smaller knots require a smaller amount of samples.
To fix the ideas, with a knot of length $L=60$, three million samples
are enough 
to evaluate $\varrho(C)$ with a variance of about $0.2$.
For a knots of length $125$, instead,
with a number of samples of five millions the obtained variance is of
the order of $1.2$.
This is not a bad result if we
consider that for a knot with $125$ segments, the volume that is needed
to be explored by the Monte Carlo sampling for the computation of
$\varrho_2(C)$ is equal to $\frac{125^4}{24}\sim 
1.0\times 10^7$. Supposing that we have a four-dimensional hypercube of such a
volume, the length of its sides will be around $56$ lattice units.
To evaluate $\varrho_2(C)$ with five millions samples means that each
dimension is explored in the 
average only
about $47$ times.
In order to tune the computational time, the two most relevant parameters
are the number of sampling points $\Upsilon$ used in the Monte Carlo
integration 
and the variance $\Sigma$. 
$\Upsilon$ and $\Sigma$ are related. If $\Upsilon$ is
too small,
the error in the estimation of $\varrho(C)$ becomes large. In that case,
the Monte Carlo
evaluation 
of the integrals contained in $\varrho(C)$ is faster, but the
rejection rate of the pivot transformations increases due to the
high uncertainty on $\varrho(C)$. On the other side, if
$\Upsilon$ is too big, the rejection rate of the pivot
transformations decreases, but the time needed for the calculation of
$\varrho(C)$ 
becomes unacceptably long.
A good choice
 is to fix $\Upsilon$ in such a way that
the variance is approximately equal
to $1$.
This is a safe estimation of the maximum possible error,
because, as mentioned before, the minimum difference between two
nearest non-coinciding values of $\varrho(C)$ is $2$.

The above evaluations of the performance of the calculations
have been made by
assuming no particular action in order to improve the computational time.
For instance, without any problem it is possible to reduce the size of
the knot by a factor three by replacing every set of three contiguous
segments with a single one. It is easy to realize that this does not
alter the knot topology on a simple cubic lattice.
Essentially, after this procedure a knot is obtained, whose length is
one third of the length of the original knot.
Of course, the new knot is defined off lattice.
Another possibility to speed up the calculations
consists in  detecting particular elements
of the knot that can be safely replaced by shorter ones.
With these tricks the problem of
studying the thermal properties of polymer knots with
length up to $L\sim 400$ becomes treatable.

\section{The Wang-Landau method}\label{sectionIV}

The Wang-Landau (WL) method \cite{wanglandau}
used here to compute the density of states has
been already extensively discussed in the physical literature. 
Its convergence has been rigorously proven in \cite{zhoubhatt}.
Here we limit ourselves to a brief review
 concerning the application of the WL algorithm to polymer
knots. 

Basically, the WL method is a self-adjusting procedure to compute the
so-called density of states $\phi_i$. For instance, let us consider the
partition function 
\begin{equation}
Z=\int {\cal D}X e^{-\beta H(X)}
\end{equation}
of a system with Hamiltonian $H(X)$, where $X$ is a
possible microstate of the system and $\beta=\frac{1}{T}$ denotes the
Boltzmann factor in thermodynamic units, in which the Boltzmann
constant is set to one: $k_B=1$.
Let us suppose that the admitted energy values $E_i$ are discrete with
$i=0,1,\ldots$, so that $Z$ may be rewritten in the form
\begin{equation}
Z=\sum_i e^{-\beta E_i} \phi_i
\end{equation}
where
\begin{equation}
\phi_i=\int {\cal D}X \delta_{E_i,H(X)}\label{dos}
\end{equation}
To compute the $\phi_i'$s, the WL algorithm proceeds as follows. Let
$g(E_i)$ denote the would be density of states and $M(E_i)$ the energy
histogram. At the zeroth approximation, we put: 
\begin{eqnarray}
g^{(0)}(E_i)=1, && M(E_i)=0
\end{eqnarray}
Successively, a Markov chain of microstates $X_{(1)},X_{(2)},X_{(3)},\ldots$
is generated. In our case, the $X_{(i)}$'s differ from each other by
transformations in which an element of the knot's trajectory of length $N$ is
changed by using the so-called pivot moves.  We use a set of three
possible pivot moves, called the inversion, reflection and interchange
transformations. They have been discussed in Ref.~\cite{pivot}, 
which the interested reader may consult for further details on this
subject. Both the location of the element of the knot to be
transformed and the kind of pivot transformation to be applied, are
randomly selected. Also the number $N$ is chosen randomly between a
given interval.

The probability of transition from a microstate $X_i$ of energy $E_i$ to a
microstate $X_{i'}$ of energy $E_{i'}$ is given by: 
\begin{equation}
p(i\rightarrow i')=\min\left[1,\frac{g^{(0)}(E_i)}{g^{(0)}(E_{i'})}\right] \label{probrule}
\end{equation}
The microstate $X_{i'}$ is accepted 
only if $p(i\rightarrow i')\ge \eta$,
$\eta$ being a randomly generated number in the interval $[0,1]$.  
If the condition  $p(i\rightarrow i')\ge \eta$ is not satisfied, the
old microstate $X_{i}$ is accepted once again.
In both cases, once a new microstate with energy  $E_{j}$ has been
selected with $j=i'$ or $j=i$, the
corresponding 
would be density of 
states $g^{(0)}(E_{j})$ and the energy histogram are updated as 
shown below: 
 \begin{eqnarray}
g^{(0)}(E_{j})&=&f_0g^{(0)}(E_{j}), \label{eqone} \\
M(E_{j})&=&M(E_{j})+1 \label{eqtwo}
\end{eqnarray}
where $f_0>1$.  Here we put $f_0=e$. We remark
that
Eq.~(\ref{eqone}) modifies the probability that microstates of energy
$E_{j}$ are accepted. In fact, the next time in which
a microstate of this kind will randomly appear after a pivot transformation,
its probability to be selected
by the
rule of Eq.~(\ref{probrule}) will be damped by a factor $f_0$. 
This
procedure of sampling new microstates that are chosen or rejected
according to the transition probability (\ref{probrule}) continues
until the energy histogram becomes flat. Since in a real simulation it
is nearly impossible to obtain a completely flat histogram,
a deviation of no more than 20\% of the $M(E_i)'$s from their average value
is admitted. Clearly, in order to have an almost flat
energy histogram, 
the microstates corresponding to different energies $E_i$ should be
almost equiprobable. In other words, after the WL procedure is
completed, the probability
of the occurrence of microstates with
energy $E_i$ is a constant independent of $i$. Let's call this
probability the WL probability and denote it with the symbol $P_{WL}(E_i)$.
To relate $P_{WL}(E_i)$ with the density of states $\phi_i$,
we remember that microstates are randomly generated 
with the help of pivot transformations.
Thus, the WL probability $P_{WL}(E_i)$ must be
equal to the unbiased probability 
$P_{unbiased}(E_i)$ of obtaining a microstate of energy $E_i$
by pivot transformations times 
the damping factor $(g^{(0)}(E_i))^{-1}$ computed
using the WL algorithm explained before. In formulas:
\begin{equation}
P_{WL}(E_i)=g^{(0)}(E_i))^{-1}\times P_{unbiased}(E_i)\label{equocan}
\end{equation}
At this point, we assume that the unbiased probability
$P_{unbiased}(E_i)$ is proportional to the density of states $\phi_i$.
More precisely:
\begin{equation}
P_{unbiased}(E_i)=\frac{\phi_i}{\sum_j\phi_j}\label{equo}
\end{equation}
where the sum over all possible energies
$\sum_j\phi_j$ is an irrelevant constant.
The above relation is intuitive, because the greater is the density of
microstates for a given value of the energy $E_i$, the greater is the
probability to obtain one of such microstates by random
transformations. Substituting Eq.~(\ref{equo}) in Eq.~(\ref{equocan}),
we arrive at the desired result:
\begin{equation}
P_{WL}(E_i)=g^{(0)}(E_i))^{-1}\times 
\frac{\phi_i}{\sum_j\phi_j}
\label{equocaneq}
\end{equation}
If the energy histogram is flat, also $P_{WL}(E_i)$ becomes a
constant, so that it is possible to write up to  irrelevant constants
\begin{equation}
\phi_i=g^{(0)}(E_i)
\end{equation}
Since the $g^{(0)}(E_i)'$s are delivered by the WL algorithm, also the
density of states $\phi_i$ is known.

Actually, if $f_0$ is too
big, the statistical errors on the $g^{(0)}(E_i)$'s may
grow large and the above equation is satisfied very roughly. 
On the other side, if $f_0$ is too small, it is necessary an enormous number of
microstates during the sampling in order to derive the $g^{(0)}(E_i)$'s.
For this reason, in the WL procedure
the density of states is computed by successive
approximations. Let us introduce to this purpose the modification
factors $f_\nu$, with $\nu=0,1,\ldots$ and $f_0=e$.
At the beginning of the $\nu-$th approximation, the value of the
factor $f_{\nu-1}$
is decreased using the relation:
\begin{equation}
f_\nu=\sqrt{f_{\nu-1}}
\end{equation}
Moreover, the would be density of states $g^{(\nu)}(E_i)$ is
initialized in such a way that it coincides with the density of states 
$g^{(\nu-1)}(E_i)$ obtained from the $(\nu-1)-$th approximation:
\begin{equation}
g^{(\nu)}(E_i)=g^{(\nu-1)}(E_i)
\end{equation}
Finally, the energy histogram is set to zero. 
At this point, the would be density of states 
$g^{(\nu)}(E_i)$ is computed at the next order by generating new
microstates and applying the same procedure used above to evaluate $g^{(0)}(E_i)$.
One should proceed in this way until, for some integer $\bar \nu$,
the modification factor $f_{\bar \nu}$ becomes sufficiently
small,  $f_{\bar\nu}\sim 1\cdot 10^{-8}$ according to the original
article \cite{wanglandau}
and the changes in the $g^{(\nu})(E_i)'$s become statistically
irrelevant.

To conclude this Section, a digression on the ergodicity
of the pivot moves used here is in order.
In the work~\cite{pivot}, this ergodicity  has been proved on a cubic
lattice for $d-$dimensional 
self avoiding walks (SAW) with fixed ends, including those in which
the ends are located at the distance of one lattice size and thus may
be considered as closed. More precisely, it has been verified in
\cite{pivot} that, starting from an arbitrary SAW of length $L$ in
$d-$dimensions, it is possible to reduce it by a finite number of
pivot moves to a given canonical SAW of the same length. Of course, no
special restriction has been required on these moves concerning their
ability to preserve the topology of a knot. Thus, some of those  moves
may allow the crossings of the lines of the SAW, a fact that can
potentially destroy the topology of the knot. For that reason, the proof
of the ergodicity of the pivot moves defined in \cite{pivot} is not
sufficient in our case, in which the topology of the studied polymer knot is
fixed since the beginning. A complete proof would require to show
that it is possible to reach,
starting from an arbitrary knot configuration, a given seed
configuration of that  knot by applying successive pivot moves.
Moreover, the knot obtained after each move should have the same topology of
the initial one. Despite the fact that we did not succeed to obtain
such a proof up to now, our empirical investigations,
performed on various knot configurations of different lengths, seem
to suggest that the pivot moves of Ref.~\cite{yzff} are ergodic also
for real polymer knots, in which the trajectories  cannot cross
themselves and thus the topology is preserved.
 Indeed,
in all analyzed cases, it has been possible to conclude that with the
pivot moves of Ref.~\cite{pivot}, together with the WL algorithm,
conformations with every 
possible number of contacts can be visited after a
sufficiently long run. A precise definition of contacts will be
provided in the next Section after introducing the potential of the
short-range forces that will take into account of
the interactions between the monomers.

\section{Thermal properties of knots}\label{sectionV}

\subsection{Observables}\label{subsectionV-1}
In this Section, we study the thermal properties of
the polymer knots $3_1$, $4_1$ and $5_1$. 
In particular,  the specific
energy $\langle E(\beta)\rangle$, the heat
capacity $C(\beta)$ and the gyration radius $\langle 
R_G^2
\rangle(\beta)$  will be computed. 
Both cases of attractive and repulsive short-range interactions, will
be considered. 
To this purpose we introduce the following potential between two
monomers:
\begin{equation}
V_{IJ}=\left\{
\begin{array}{rl}
+\infty&\mbox{if $I=J$}\\
 \varepsilon&\mbox{if $d=|\vec {R}_I-\vec {R}_J|=1$ and
  $I\ne J\pm 
  1$}\\
0&\mbox{otherwise}
\end{array}
\right.\label{potdef}
\end{equation}
with $\varepsilon$ being a small energy scale determining the
strength
interactions between pairs of non-bonded
monomers. The condition $\varepsilon< 0$ characterizes the  attractive
case, while $\varepsilon>0$
characterizes the repulsive case.
Moreover, $\vec{R}_{I}$ denotes the
position vector of the $I$-th segment.
Conformations such that $I=J$ are automatically discarded, because no
crossing of the trajectories is possible in a real polymer knot.
Besides, when a crossing occurs, the knot is no longer mathematically well
defined. 

The total Hamiltonian of a polymer knot in a given
microstate $X$ is given by:
\begin{equation}
H(X)=\sum_{I,J=1}^LV_{IJ}
\end{equation}
To simplify the expression of $H(X)$, it is convenient to classify the
microstates according to their number of contacts $m$.
Two non-bonded monomers are said to
form a contact
if their reciprocal distance on the lattice is equal to $1$ (see the
second condition in Eq.~(\ref{potdef})).
$m$ counts the number of contacts of every pair of noncontiguous monomers
appearing in the conformation.
Clearly, for a microstate $X_m$ with a number $m$ of contacts and with
no overlapping monomers, the
Hamiltonian reads as follows:
\begin{equation}
H(X_m)=m\varepsilon
\end{equation}
The density of states $\phi_{m}$
defined by Eq.~(\ref{dos})
is computed by means of the Wang-Landau algorithm illustrated in Section
\ref{sectionIV}.  

In our settings, $\langle
E(\beta)\rangle$ 
and $C(\beta)$
are
expressed as follows: 
\begin{eqnarray}
\langle E(\beta)\rangle &=&\dfrac{\sum_{m}m\varepsilon e^{-\beta m
    \varepsilon}\phi_{m}}{\sum_{m} e^{-\beta m
    \varepsilon}\phi_{m}} 
\label{energy}\\
 C(\beta)&=&\dfrac{1}{T^{2}}(\langle E(\beta)^{2}\rangle -\langle
 E(\beta)\rangle^2)\label{heatcapacity}
\end{eqnarray}
while
the mean square radius of gyration $\langle 
R_G^2
\rangle(\beta)$
is given by \cite{yf,pnv}:
\begin{equation}
\langle R_{G}^2
\rangle (\beta)  =\dfrac{\sum_{m}{\langle
    R_{G}^{2}\rangle}_{m}  e^{-\beta m 
    \varepsilon}\phi_{m}}{\sum_{m} e^{-\beta m
    \varepsilon}\phi_{m}} 
\label{rg}
\end{equation}
with $\langle R_{G}^{2}\rangle_{m}= \frac 1{2L^2}\sum_{I,J=1}^{L}
\langle(\vec {R}_I-\vec{R}_J)^{2}
\rangle_m$ denoting the average of the gyration radius
computed over states with $m$ contacts. 

Finally, we wish to  discuss how the problem of rare events 
has been treated in this work.
While there
is enough evidence that
the pivot algorithm is ergodic, it is also true that certain states
occur very rarely and sometimes may require the computations of
billions of trial conformations before being obtained. This is the case of
very compact ($m>L$) or 
very swollen ($m\sim 0$)
conformations.  
For polymers of any length $L$, at least up to $L=300$, the maximal
length that has been studied with the present approach, states with a
number of contacts higher 
than $L$ should be considered as rare events. 
For short polymer knots,
samples with $m=L$ or more are extremely rare. When $L=50$ or $70$,
the lowest energy states that have been reached after a few millions of
samples are characterized respectively by $48$ and $69$ contacts.
Short polymers which are heavily knotted are also difficult to be
found in conformations with a small number of contacts. For a trefoil
with $L=50$ or $L=70$
the state with $m=0$ should be considered as rare, while for $8_1$
$m<5$ is rare.
The problem is that rare events act like bottlenecks in the WL
algorithm, which prevent the energy histogram to become flat unless a
prohibitively high number of samples is used. Such bottlenecks
appearing in polymer simulations have been discussed in
\cite{binder}. In that reference, the introduction of a cutoff in the
energy and a slightly different criterion of flatness has been proposed.
Accordingly, our simulations have been
restricted to conformations in which the number of contacts 
is limited by the condition
$m\le L$. For short and topologically complex knots, $m$ has additionally
been bounded from below to be greater than $0$. In the case of some
knots, 
which are not
considered here, the
restriction $m<5$ has been required. 
These cuts in the number of contacts $m$ affect the calculations
at lower and higher temperatures at most by a relative error of a few
percent. The 
data about the average gyration radius show for example that the
difference between the mean square gyration radii obtained 
by cutting or not the five lowest values of $m$ is very small.
On the other side, without imposing bounds on $m$, we have
observed that the flattening procedure of the energy histogram in the 
Wang-Landau algorithm requires an enormous amount of sampling,
especially if the rare events occur when the modification factor
$f_\nu$ is small.

\subsection{Simulation results based on the knot invariant $\varrho(C)$}
\label{subsectionV-2}
\begin{figure*}
\centering
\includegraphics[width=0.50\textwidth]{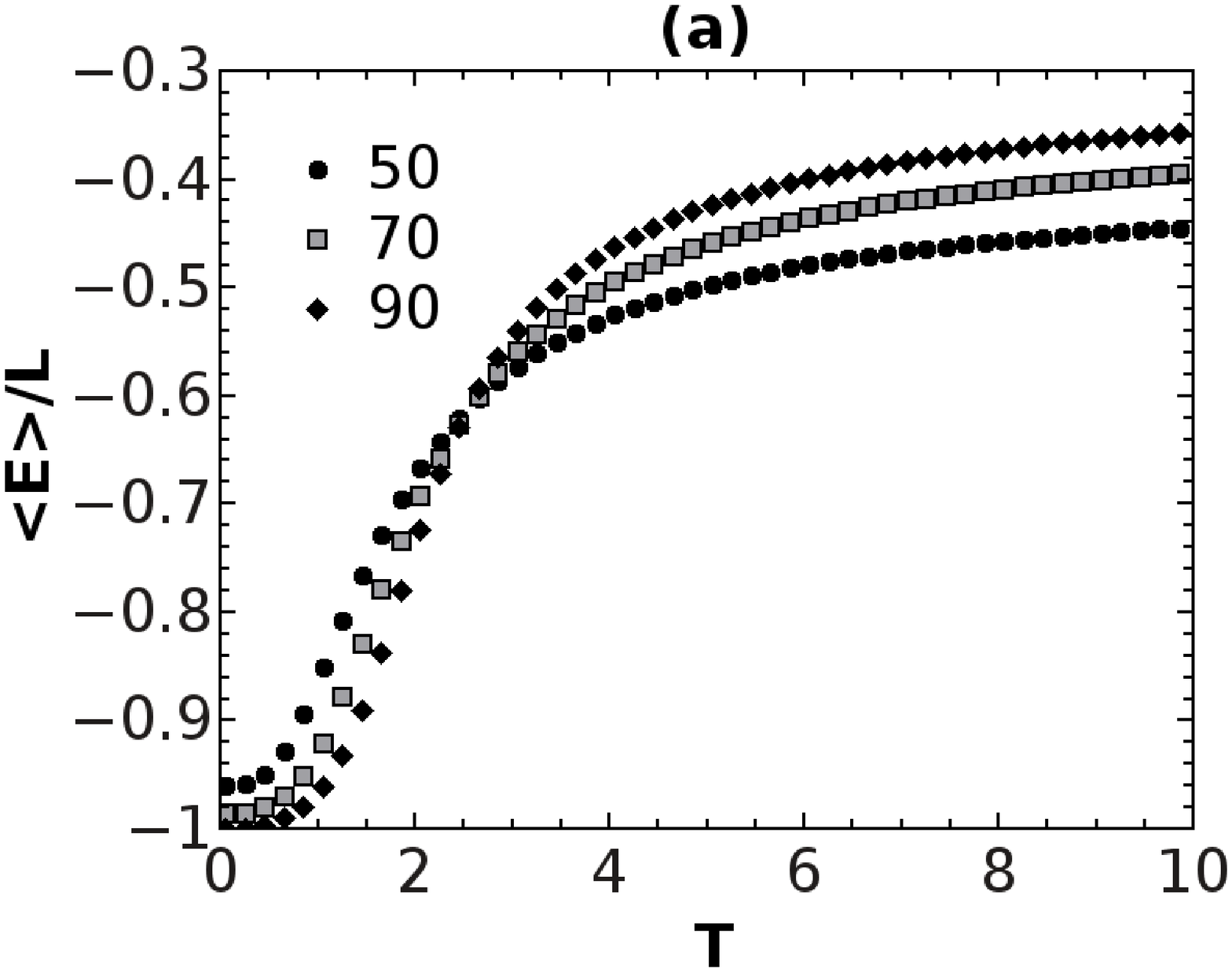}\includegraphics[width=0.50\textwidth]{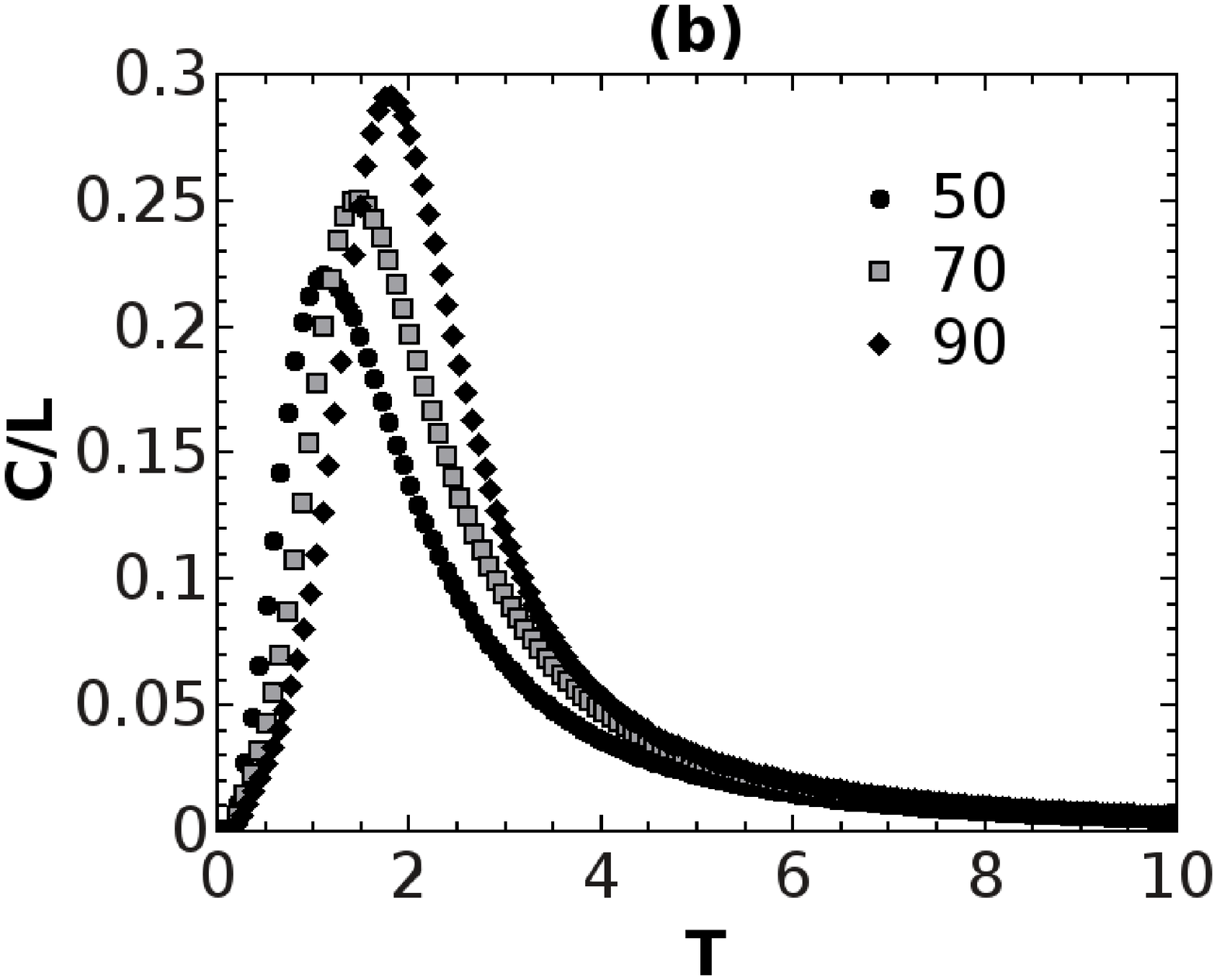}
\caption{\label{fig1}  In the attractive case, the specific energy (in units of $\varepsilon$) and
  heat capacity of the trefoil as functions of the normalized
  temperature $\mathbf T=\frac{T}{\varepsilon}$. The polymer length can
  take the values $L=50$ (circles), $70$ (rectangles) and $90$ (diamonds).} 
\end{figure*}
In Fig.~\ref{fig1}, left panel, the results for the specific energy 
 and the heat capacity in the attractive case are displayed. 
Let us note that in all figures we have used the symbol $\mathbf T$
for the normalized temperature $\frac T\varepsilon$.
As it is
 possible to see, the specific energy
increases with increasing
temperatures as it should be, because as the temperature grows, more
and 
more energetic states are excited. Moreover, at high temperatures
longer polymers have a higher
specific energy than shorter ones.
This behavior may be explained by the following two observations.
The first observation, which is true for both short and long polymers,
is that, 
when the temperature is
very high,
the energy gap
$\varepsilon<0$ of the potential $V_{IJ}$ defined in
Eq.~(\ref{potdef}) is much smaller than the
energy related to the thermal fluctuations. 
 As a consequence, 
the
polymer is supposed to be in a more swollen conformation than at
low temperatures, where the interactions dominate over the thermal
fluctuations. 
The second observation is that the effects of knotting are less and
less important with increasing polymer lengths. 
To convince ourselves that this is the case, it
is sufficient to imagine a knot which is localized in a small part of
the polymer. If the polymer is long, the localized part will be not
relevant in comparison with the  rest of the polymer, which will
behave more or less like a unknotted ring. 
Also numerical simulations confirm this trend, see for example
\cite{yzff}. Moreover, our calculations show that
the differences in the specific energy and heat capacity between two
trefoils of lengths $L=90$ and $L=120$ are much less marked
than the same differences between  two trefoils of lengths $L=70$ and
$L=90$. Indeed, the data of the trefoil with $L=120$ have not been
reported in Fig.~\ref{fig1}(a) and (b) because it is difficult to distinguish
them from those of the trefoil with $L=90$.
Combining the two observations above, one can conclude that, at high
temperatures, the polymer will tend to swell, but the effect of the
topological constraints  
will be to counteract this swelling process.
On the other side, the topological constraints and their effects
become less important when polymers are long. As an upshot, if the
temperature $\mathbf T$ and the knot topology are kept fixed, we expect that 
the average distance between the monomers and thus the specific energy
of the knot will increase with increasing polymer length when $\mathbf T>>1$.
This explains the behavior of the specific energy of Fig.~\ref{fig1}
at higher temperatures.
The behavior of $\langle E(\beta)\rangle$ at lower temperatures should
be taken with some care because, as already mentioned, the number of contacts
has been limited by the condition $m\le L$, so that the lowest energy
state cannot be reached. 
Always for that reason, the specific energy at zero temperature is
slightly different 
for different lengths.

Concerning the heat
capacity, we can see in Fig.~\ref{fig1}, right panel,
 that there is only one sharp peak in
the whole temperature region.
The interpretation of this peak is somewhat difficult.
It is certainly a
pseudo phase
transition caused by the fact that we are working on a lattice with a
finite size system, as those discussed in
Refs.~\cite{wuest,janke}.   
Similar pseudo phase transitions have been
already observed in knots, see \cite{swetnam}. 
In \cite{janke} it has been stressed that, along with the advances in
constructing high resolution equipment, such pseudo phase transition
in real system become more and more important.
It is likely that the peaks in the heat capacity
correspond to the transition of the
knot from a frozen crystallite state to an expanded state
similar to what
happens in the case of a single polymer chain discussed
in~\cite{binder}. The data concerning the gyration radius in
Fig.~\ref{fig3}, left panel, seems to confirm this hypothesis, because
the gyration radius stats to grow abruptly more or less in the same range of
temperatures in which the peak is appearing.
This hypothesis could also explain the absence of a second peak or shoulder
connected to the coil-globule transition. As it was discussed in
\cite{binder}, in fact, if the range of the interactions is very short,
then an open chain admits just two possible states, the crystallite and
the expanded coil ones. For a knot, the situation is however more
complicated.
The visual analysis of the samples shows that the lowest energy
conformations exhibit some kind of ordering, but probably a full
ordering is forbidden by the topology of the knot.
In conclusion, further analysis is necessary in order to identify
the nature of these transitions.
\begin{figure*}
\centering
\includegraphics[width=0.5\textwidth]{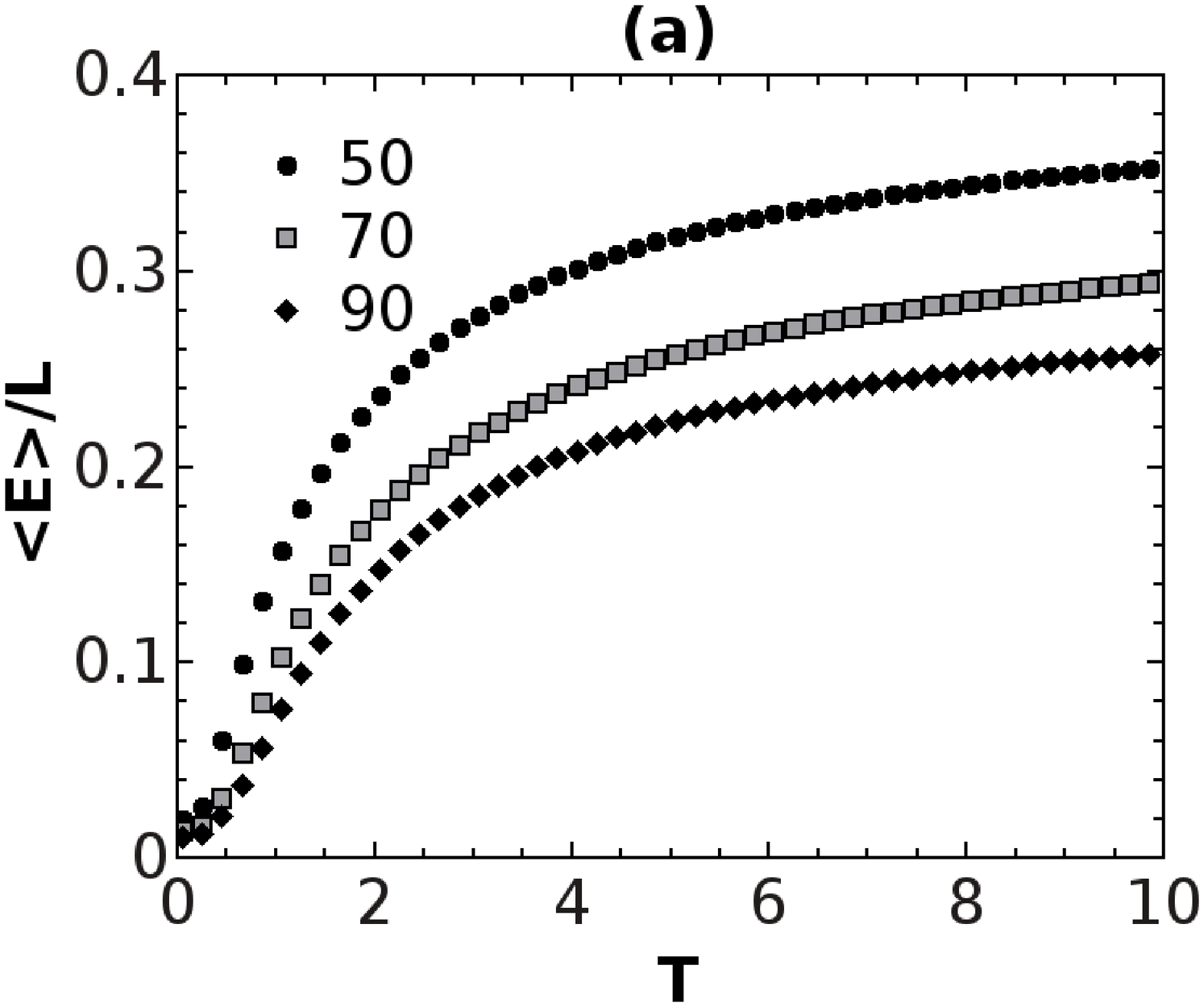}\includegraphics[width=0.5\textwidth]{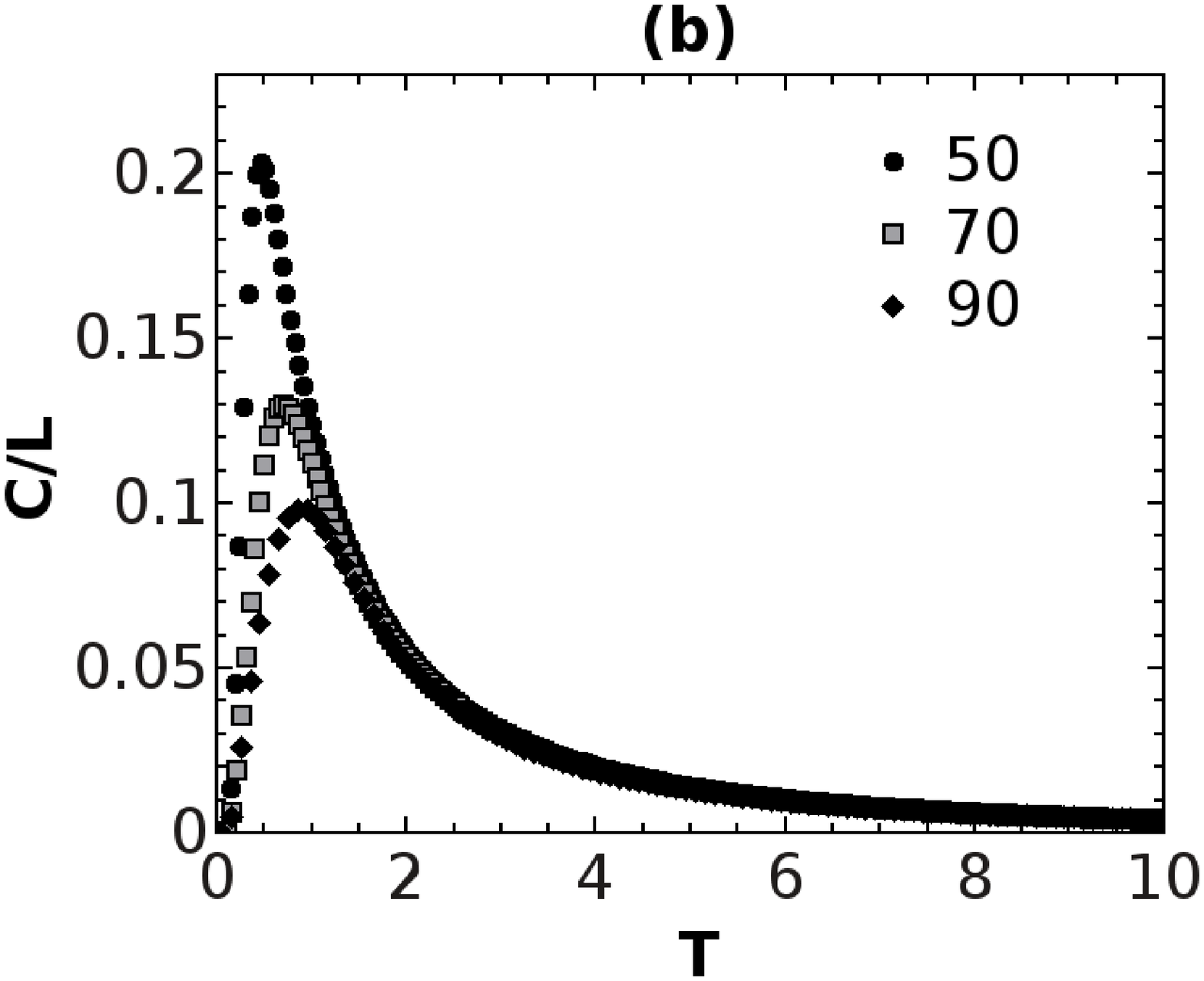}
\caption{\label{fig1-r}  In the repulsive case, the specific energy (in units of $\varepsilon$) and
  heat capacity of the trefoil as functions of the normalized
  temperature $\mathbf T=\frac{T}{\varepsilon}$. The polymer length can
  take the values $L=50$ (circles), $70$ (rectangles) and $90$ (diamonds).} 
\end{figure*}

Analogous considerations can also be made in the repulsive case displayed
in Fig.~\ref{fig1-r}. 
One difference is that low temperatures
correspond now to the swollen state, while at high temperatures, when
the energy barrier $\varepsilon$ becomes negligible with respect to
the energy carried by the thermal fluctuations, the monomers are
allowed to get nearer and the average number of contacts is growing. 
As a consequence,
the gyration radius for repulsive interactions decreases at higher
temperatures
as shown in
Figs.~\ref{fig3}(b) and \ref{fig3b}(b). Moreover, while as expected
 the specific energy increases with increasing temperatures,
it turns out that, contrarily to what happens when the forces are
attractive, longer polymers have a lower
specific energy than shorter ones at any temperature.
To explain this decrease of the energy of longer polymers, we recall that
the swelling of
the polymer knot is hindered by the topological constraints, in such a
way that more complex knot configurations correspond to more compact
polymer conformations. 
As well, due to the fact that those parts of the polymer trajectory which
are affected by the  
topological constraints will become less and less important when
the polymer length increases, it is licit to conclude that longer
polymers
will in the average admit conformations that are more swollen with
respect to those of shorter polymers as we argued in the attractive case.
The difference is that, if the interactions are repulsive, more 
swollen conformations are less energetic, which explains why the
specific energy of longer polymers is in the average lower 
than that of shorter polymers.
The presence of sharp peaks in the heat capacities, see Figs.~\ref{fig1-r}(b)
and~\ref{fig2-a}(b), has not a straightforward interpretation like
those occurring when the interactions are attractive.
Apparently, as argued in \cite{yzff}, 
at $T=0$ the knot is in one of the lowest energy conformations which
are allowed. As the temperature increases, at the beginning the system
is unable to pass to a more compact configuration unless
$T\sim\varepsilon$. After that threshold, the knot more and more
contacts are possible between the monomers unless saturation is reached.

\begin{figure*}
\centering
\includegraphics[width=0.5\textwidth]{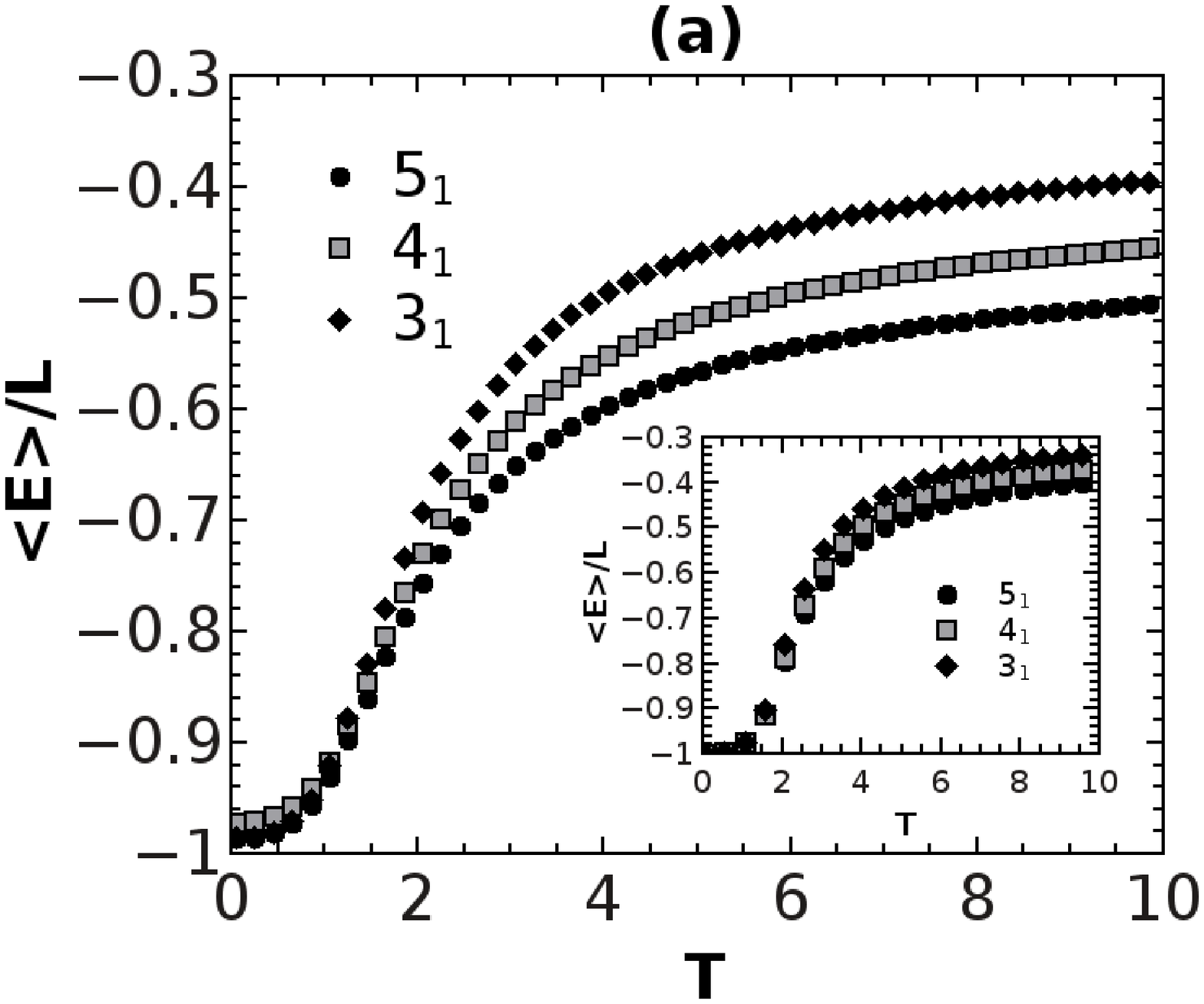}\includegraphics[width=0.5\textwidth]{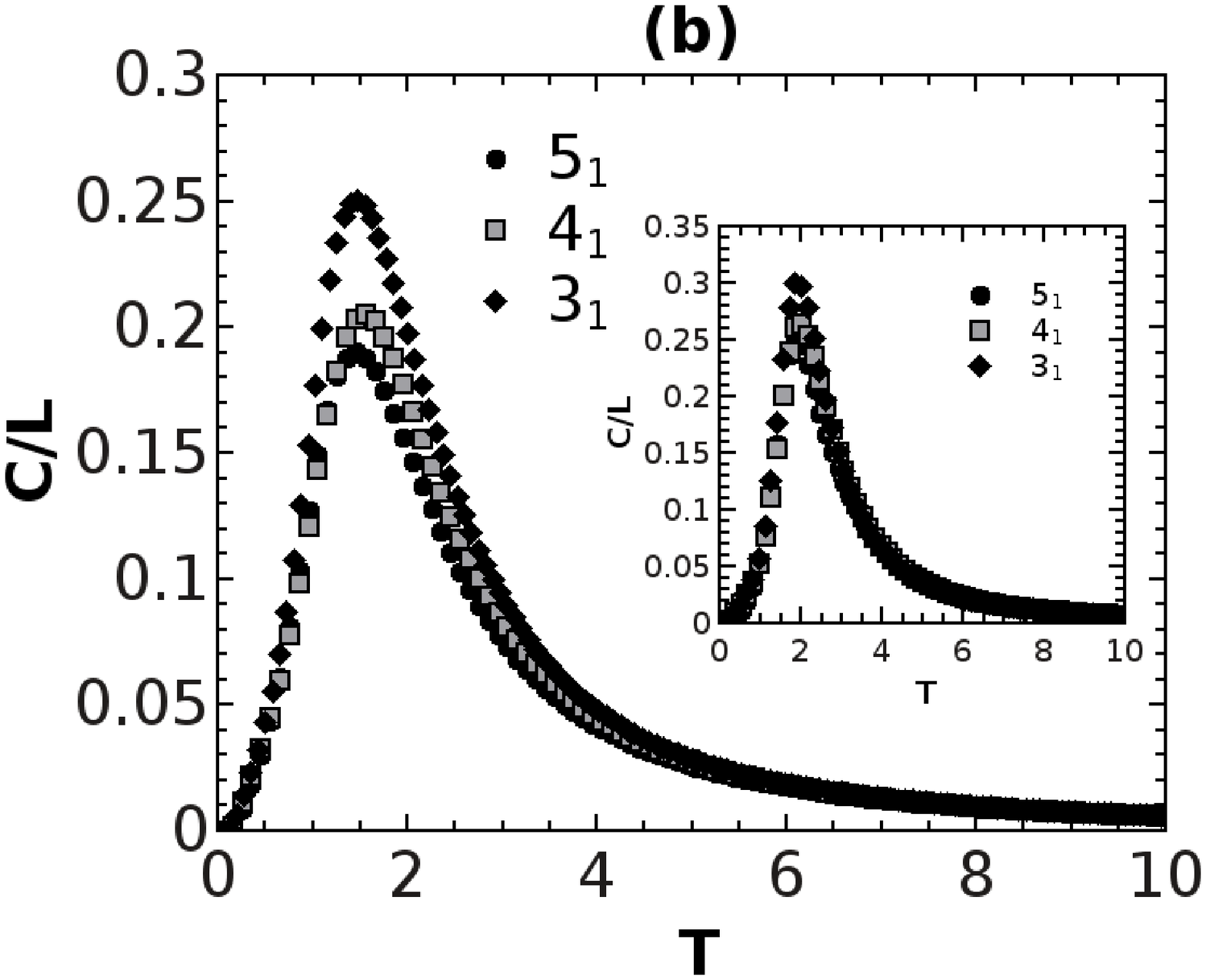}
\caption{\label{fig2}  Specific energy (in units of
  $\varepsilon$) and heat capacity for knots $5_1$, $4_1$ and $3_1$ in the
  attractive case. The polymers have length $L=70$. Inset: Specific energy and heat capacity for knots $5_1$, $4_1$ and $3_1$ with length $L=120$ in the
  attractive case.} 
\end{figure*}

\begin{figure*}
\centering
\includegraphics[width=0.5\textwidth]{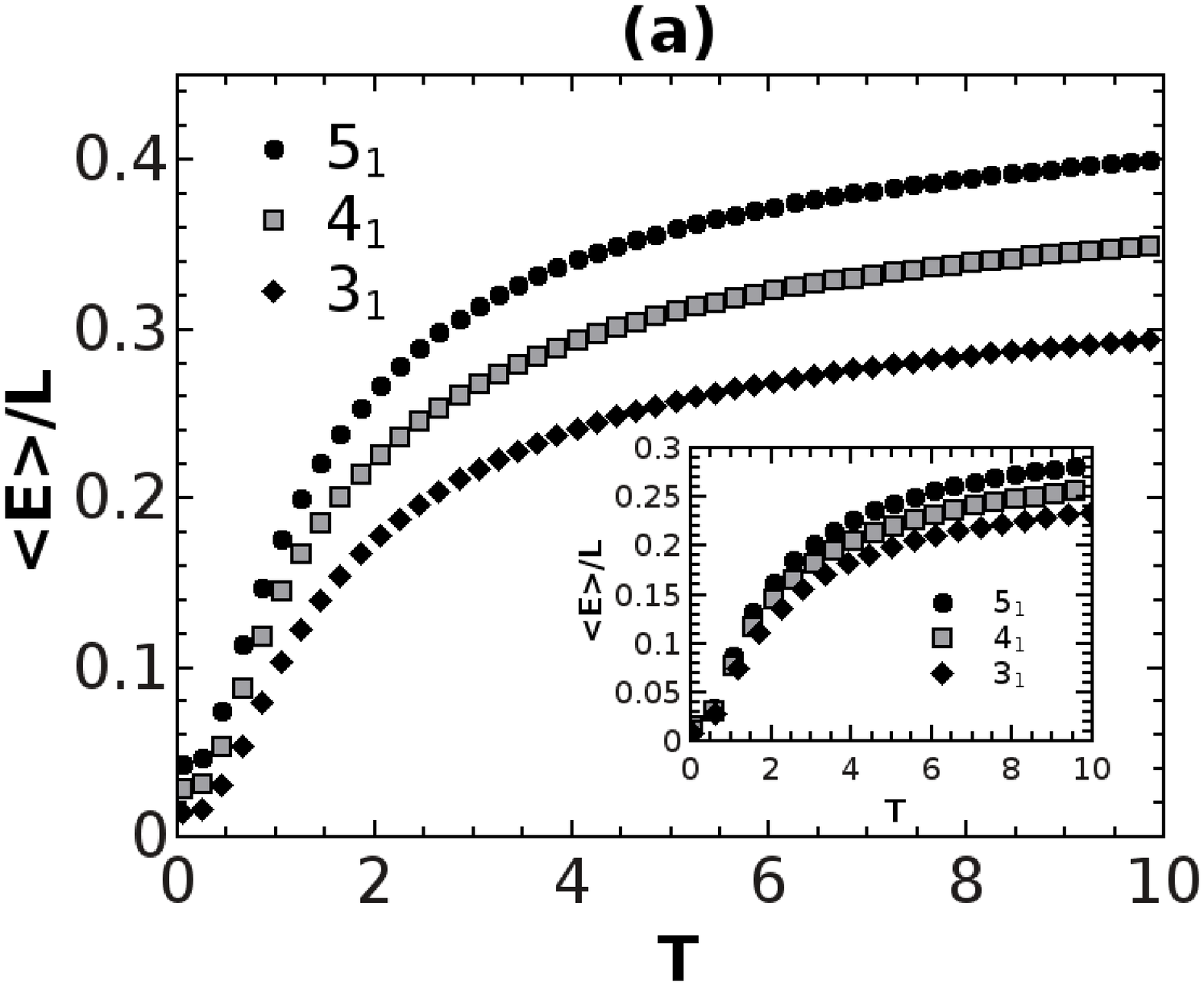}\includegraphics[width=0.5\textwidth]{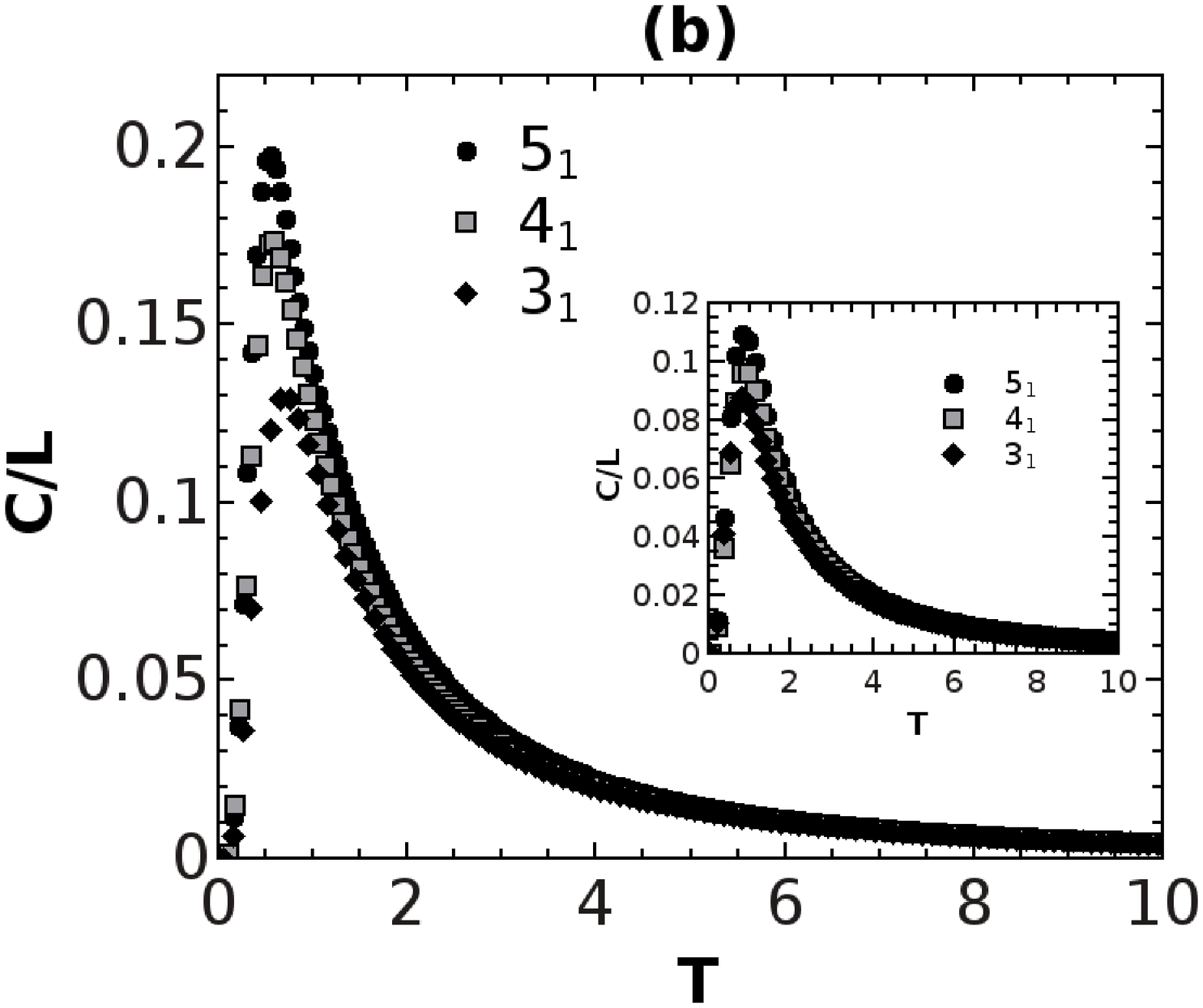}
\caption{\label{fig2-a}  Specific energy (in units of
  $\varepsilon$) and heat capacity for knots $5_1$, $4_1$ and $3_1$ in the
  repulsive case. The polymers have length $L=70$. Inset: Specific energy and heat capacity for knots $5_1$, $4_1$ and $3_1$ with length $L=120$ in the
  repulsive case.} 
\end{figure*}

\begin{figure*}
\centering
\includegraphics[width=0.5\textwidth]{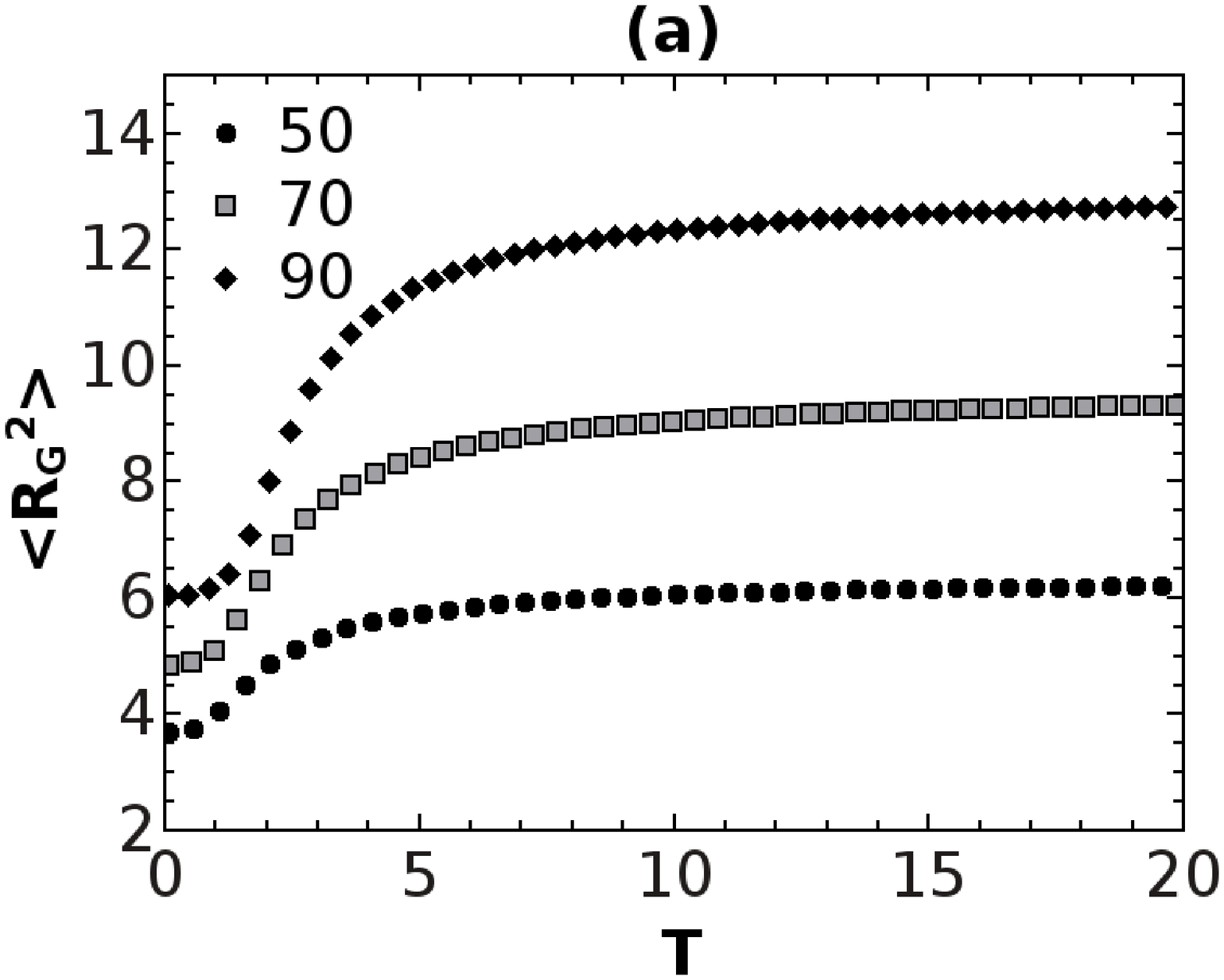}\includegraphics[width=0.5\textwidth]{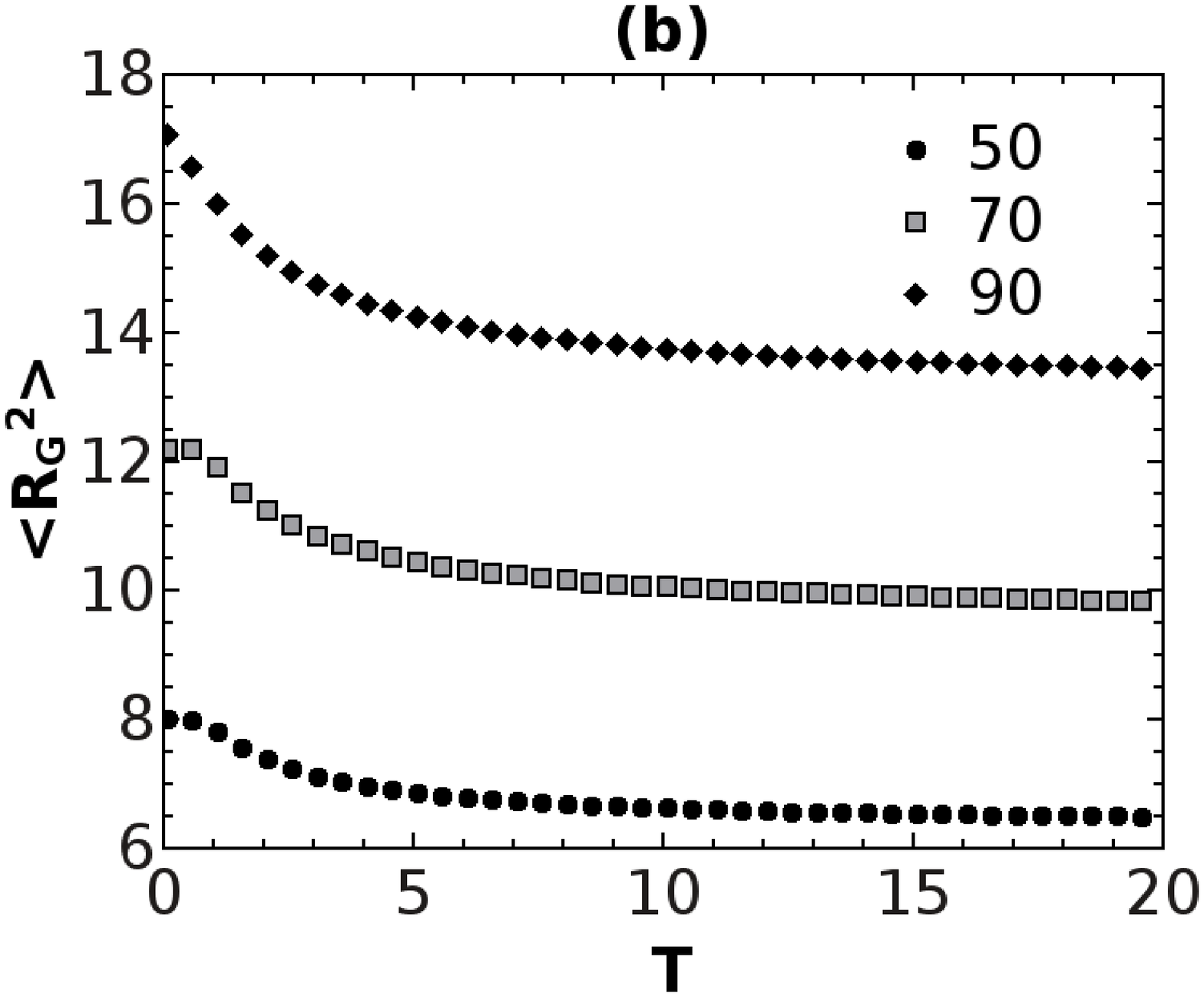}
\caption{\label{fig3}  Mean square gyration radius of trefoil knots of lengths $L=50,70,90$ as a function
  of the temperature. (a) Plot of the mean square gyration radius
  in the attractive case; (b) Plot of the mean square gyration radius in the repulsive case.
  } 
\end{figure*}

\begin{figure*}
\centering
\includegraphics[width=0.5\textwidth]{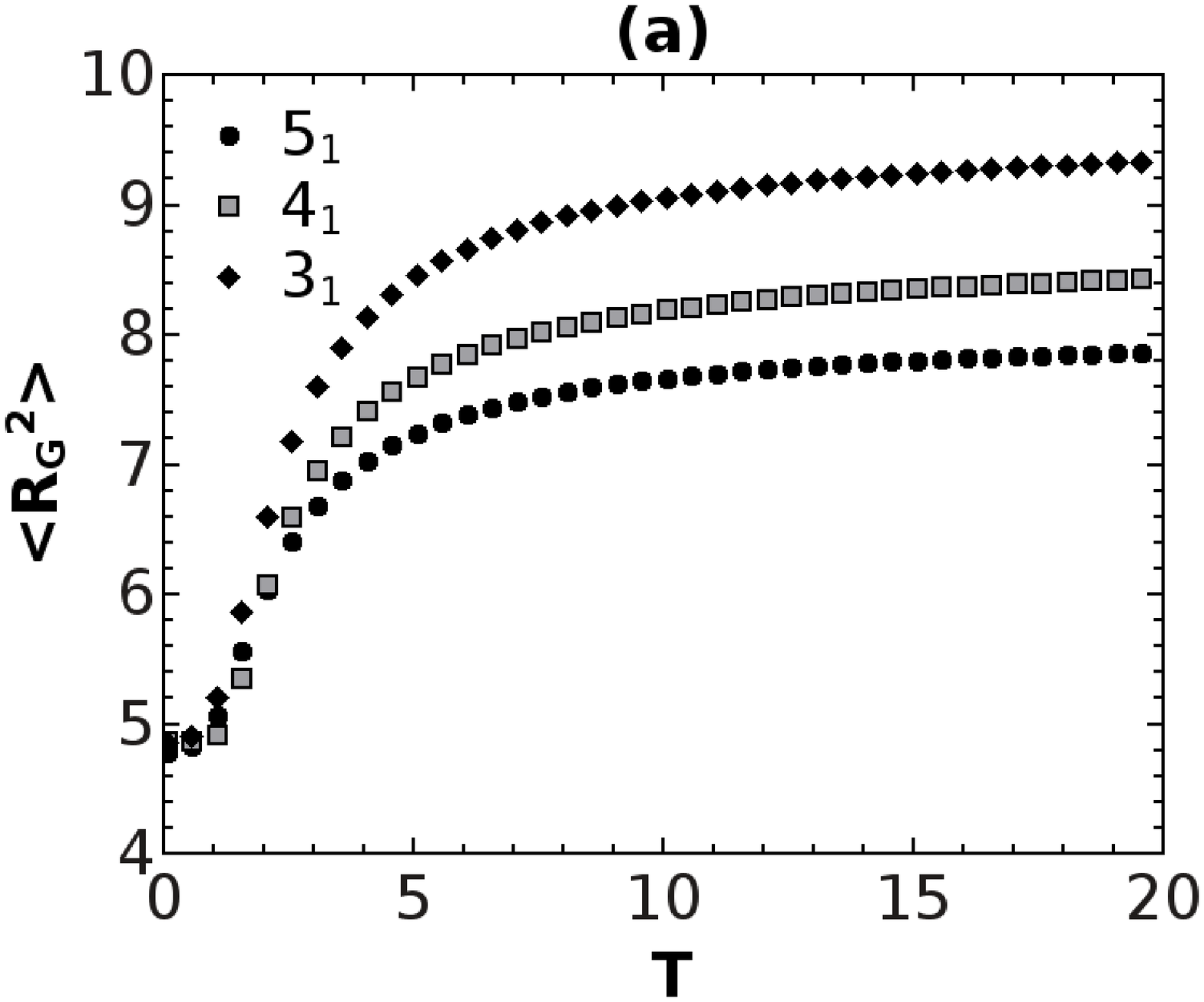}\includegraphics[width=0.5\textwidth]{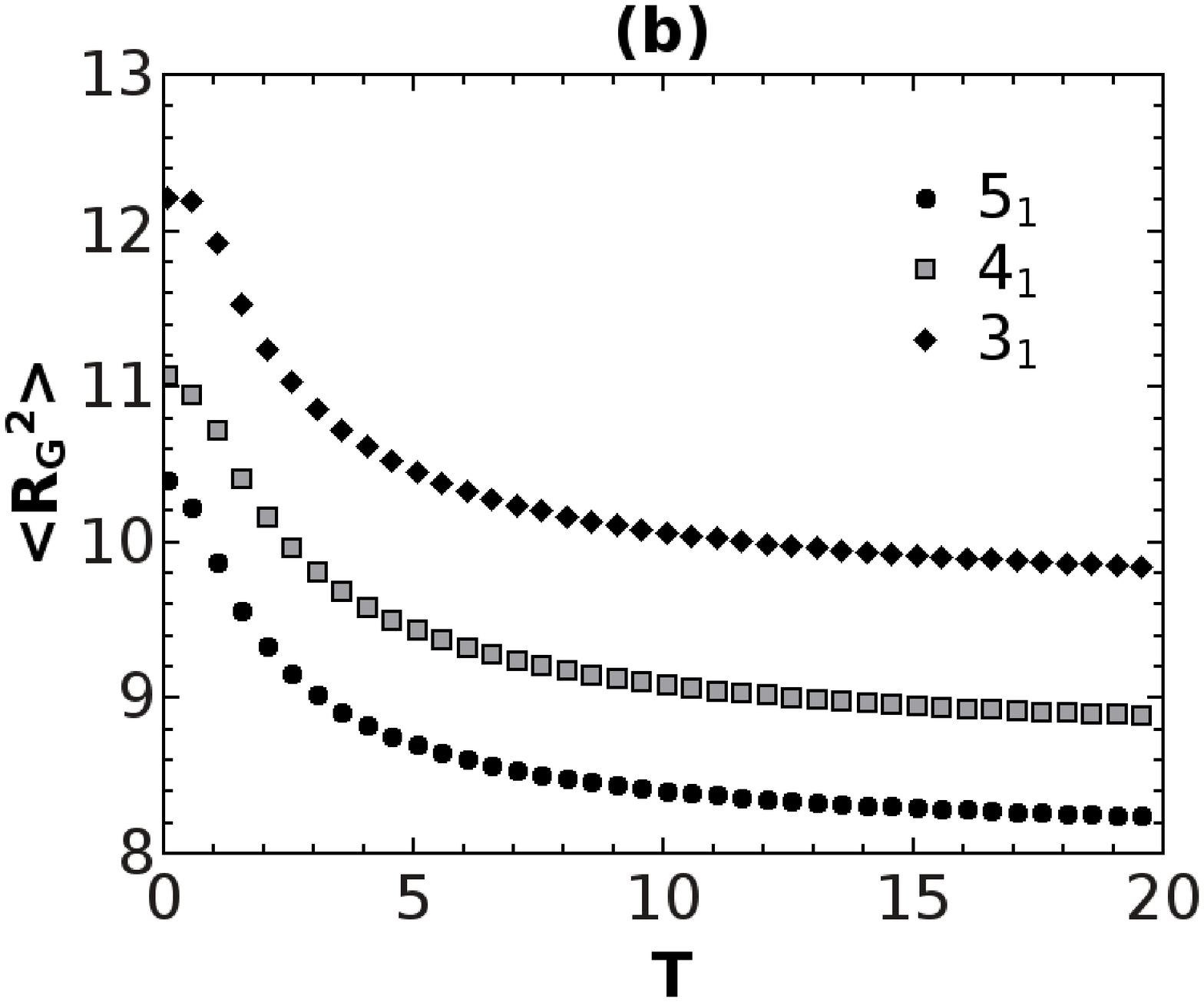}
\caption{\label{fig3b}  Mean square gyration radius of knots $5_1$, $4_1$ and $3_1$ of length $L=70$ as a function
  of the temperature. (a) Plot of the mean square gyration radius
 in the attractive case; (b) 
  Plot of the mean square gyration radius in
  the repulsive case.} 
\end{figure*}

To check the
effects of the topology on the behavior of the polymer,
we have tested different knot configurations of the same length.
We present here the data of polymers with $L=70$.
We observe that, in 
the attractive case, the increasing of the knot complexity results
in the decreasing of the specific energy and heat capacity at high
temperatures, see Figs.~\ref{fig2}(a) and (b).
As one may expect from the previous discussion, in the repulsive case
it is exactly the converse, i.~e. 
the increasing of the knot complexity results in the increasing 
of the specific energy
and heat capacity when the temperature grows, as shown in
Fig~\ref{fig2-a}(a) and (b). 
Once again, this is due to the fact that, if the topology of the knot
is more complex, the knot conformation contains more contacts
and is more compact. To have more contacts implies that, in the
average, the 
energy of each monomer and thus the specific energy is lower if the forces
are attractive and higher if the forces
are repulsive.
This scenario is confirmed by the data on the gyration radius of
different knots with length $L=70$ in both the attractive and
repulsive case, see Fig.~\ref{fig3b}.
The gyration radii $\langle R_G^2\rangle$ of the three different knots
$3_1,4_1$ and $5_1$ satisfy the inequality
$(\langle R_G^2\rangle)_{3_1}> (\langle R_G^2\rangle)_{4_1}>( \langle
R_G^2\rangle)_{5_1}$ independently if the 
interactions are attractive or repulsive.
\begin{figure}
\begin{center}
\includegraphics[width=3in]{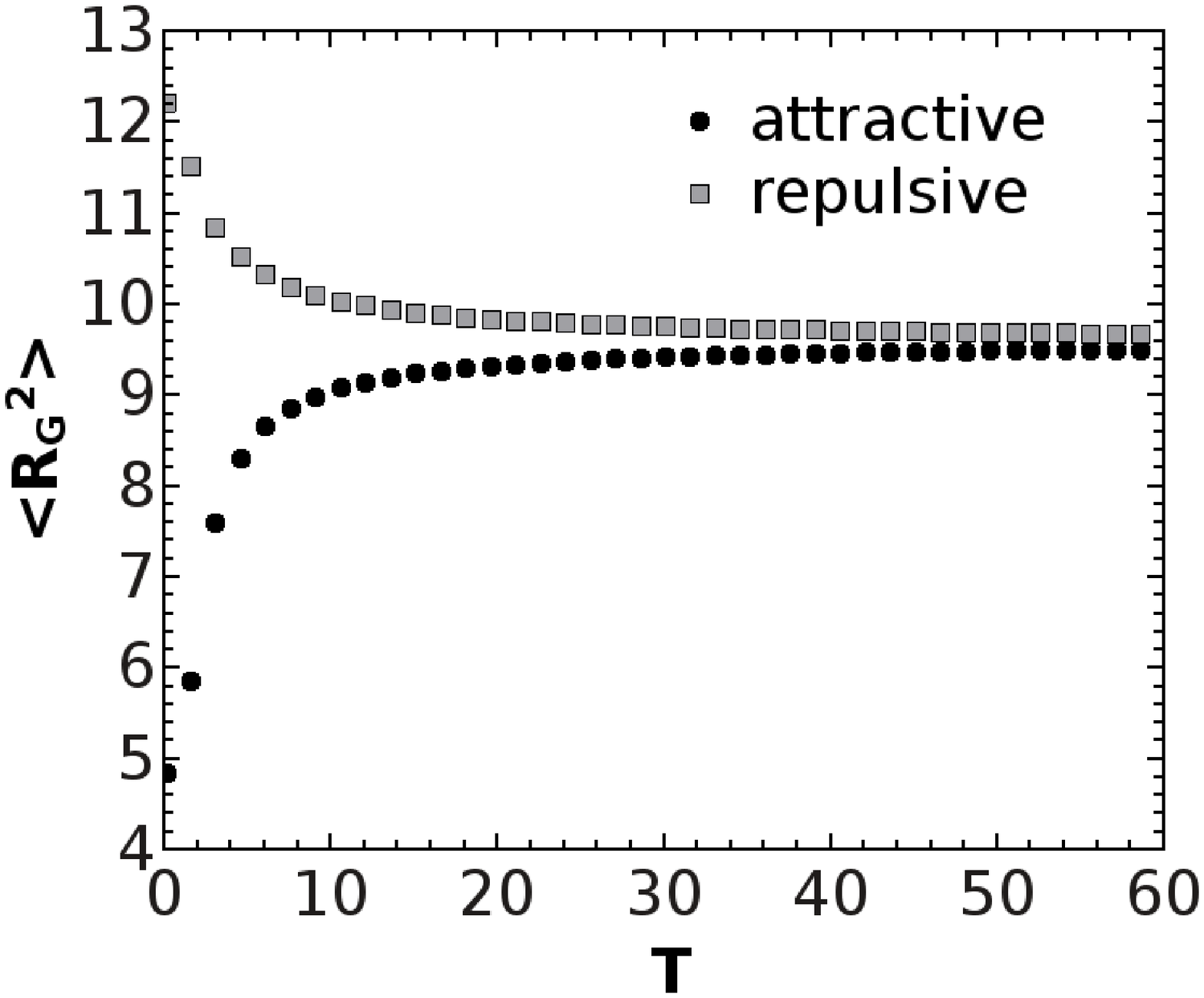}
\caption{\label{fig4}
Convergence at high temperatures of the mean square gyration radii computed in the attractive and repulsive cases for a trefoil knot of length $L=70$.} 
\end{center}
\end{figure}
Finally, as we previously mentioned, at higher temperatures the
effects of the interactions should become negligible.
Accordingly, even if the gyration radius exhibits a different
behavior depending
if the interactions are
repulsive or attractive, in Fig.~\ref{fig4} we see that the
values of  the radius of gyration computed in these two different
cases get closer and closer when the temperature is 
increasing. Eventually, they should converge if the temperature is
high  enough.  

\subsection{Comparison with the PAEA method}\label{subsectionV-3}
The PAEA method allows to use topology changing pivot transformations
in numerical simulations of polymer systems and prevents the
accidental transition to another topology without making use of
topological invariants.
The idea of the PAEA method is based on the following observation.
After a pivot
transformation is performed on a randomly chosen element $\Delta k$
of length $N$ of a knot,
$\Delta k$ is
transformed into the new element  $\Delta k'$. Since  $\Delta k$ and
$\Delta k'$ have their two ends in common, because these ends are untouched
by the pivot transformation, it is easy to realize that, together,
$\Delta k$ and $\Delta k'$ form a small loop or, if $N$ is large, a
set of small loops of total length $2N$.
The topology is preserved by checking whether
the part of the knot unaffected by the pivot transformation
crosses an arbitrary surface spanned around the small loop(s). If the
crossing happens, then the trial conformation will
be rejected and the knot undergoes another pivot
transformation. Otherwise, the trial conformation is accepted
and a new pivot transformation is applied to it.  
More details can be found in \cite{yzff}.

What is crucial for the success of the PAEA method is 
to classify all the possible small loops of $2N$ segments that can be
produced after a random pivot transformation and to construct
suitable
surfaces having these small loops as borders. 
When the length $N$ of the segments changed by the pivot
transformations in greater than five, the number
of all loops of this kind becomes large.
With increasing values of $N$, it becomes
more and more difficult to construct the surfaces mentioned above.
Unfortunately, this is a not problem that
can be solved automatically by the computer using some algorithm.
For this reason, up to now the PAEA algorithm has been developed only in
the case $N=4,5$.
The advantage of the technique discussed in this work based on the
knot
invariant $\varrho(C)$ is that the
number of segments involved in the pivot transformations may
arbitrarily range within the interval $1< N\le L$.
Larger values of $N$ change bigger portions of the knot and this leads to a
faster equilibration process than the PAEA method.
Moreover, we have observed that, as
the length $N$ of the segments to be changed is increasing, the
number of samples $N_{samples}$ needed to get a
flat energy histogram is  decreasing.
This fact is shown  in Table~\ref{f-paea-mc}, where as an example the
ratio $(N_{samples})_{PAEA}/(N_{samples})_{\varrho(C)}$ is displayed
in  the case of the knot $3_1$ with length $L=120$. 
In the calculations with the PAEA method the pivot transformations
were limited to $N=4$, while in the calculations with the $\varrho(C)$
invariant $N=24$. 
All the approximation degrees $\nu=0,\ldots,16$ used in computing the
density of states have been listed.
As we can see from Table~\ref{f-paea-mc}, the
 number of samples $(N_{samples})_{PAEA}$ needed in the PAEA method 
to make the energy histogram flat 
is always
larger than the corresponding number of samples $(N_{samples})_{\varrho(C)}$ necessary
to distinguish the topology with the help of $\varrho(C)$.
The explanation of this increasing in the efficiency of the method
based on the knot invariant $\varrho(C)$ with respect to the PAEA
method 
is that with the invariant $\varrho(C)$ large pivot transformations
are allowed and they are able to modify  a relevant portion of the polymer.
The larger is the
number $N$ of
segments affected by a pivot transformation, the greater
is the difference between the numbers of contacts of the knot before
and after the transformation.
As a consequence, 
with a large pivot transformation it is possible to jump from
conformations that have very different values of $m$, thus
accelerating considerably the exploration of the set of all possible
 conformations of a knot
compatible 
with the given topological configuration.
\begin{table}[ht]
\caption{ The ratio $(N_{samples})_{PAEA}/(N_{samples})_{\varrho(C)}$ (denoted
with $PAEA/\varrho(C)$ in this table) obtained from the computations
of
the density of states of the knot $3_1$ with length
$L=120$. The initial value of the modification factor is $f_0=e$.}  
\centering
\begin{tabular}{c c c c}
\hline\hline
 $f_\nu$ & $PAEA/\varrho(C)$ & $f_\nu$ & $PAEA/\varrho(C)$\\ [0.3ex]
\hline
$f_0=e$ &7   & $f_{9}=\sqrt{f_8}$ & 6 \\
$f_1=\sqrt{f_0}$ &4   & $f_{10}=\sqrt{f_9}$ & 4 \\
$f_2=\sqrt{f_1}$ &4   & $f_{11}=\sqrt{f_{10}}$ & 5 \\
$f_3=\sqrt{f_2}$ &7   & $f_{12}=\sqrt{f_{11}}$ & 3 \\
$f_4=\sqrt{f_3}$ &5   & $f_{13}=\sqrt{f_{12}}$ & 10 \\
$f_5=\sqrt{f_4}$ &5   & $f_{14}=\sqrt{f_{13}}$ & 5 \\
$f_6=\sqrt{f_5}$ &4   & $f_{15}=\sqrt{f_{14}}$ & 6 \\
$f_7=\sqrt{f_6}$ &4   & $f_{16}=\sqrt{f_{15}}$ & 7 \\
$f_8=\sqrt{f_7}$ &5   & & \\ 
\hline
\end{tabular}
\label{f-paea-mc}
\end{table} 
Of course, the calculation of $\varrho(C)$ implies the evaluation of
quadruple integrals with a good precision, otherwise there will be not 
a sufficient control of the topology. To perform the necessary
integrations
becomes challenging with increasing polymer lengths even using a Monte
Carlo integration method as we did here. However, since what we need
to calculate 
is the second Conway coefficient and its value is dependent on
the topological configuration of the knot, but not on
its length, it is always possible to shorten the knot
provided its topology is not affected. A simple way to do that is to
group together triplets of contiguous segments of the knot. This allows
to decrease by a factor three the length of the polymer which, after
this procedure, will 
be of course no longer defined on a simple cubic lattice.  
Such reduction method can be combined with other 
 reduction algorithms, like for instance the KMT reduction scheme
 proposed in \cite{Muthukumar,Taylor}.

\begin{figure}
\begin{center}
\includegraphics[width=3in]{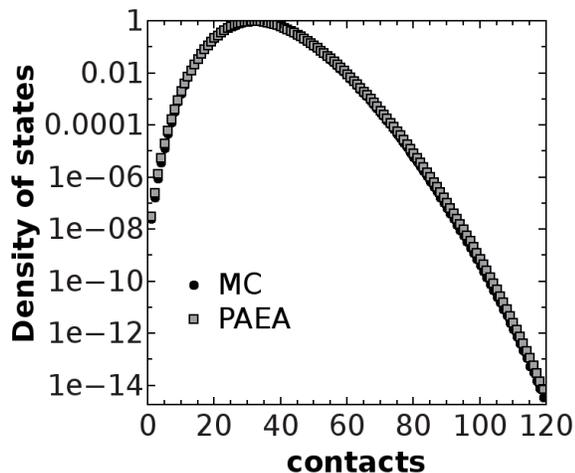}
\caption{\label{dost50}
Comparison of the densities of states $\phi_m$ for a trefoil knot with length $L=120$ computed using the PAEA method (grey squares) and the knot invariant
$\varrho(C)$ evaluated with a Monte Carlo (MC) algorithm (black
circles).
} 
\end{center}
\end{figure}

Finally, we wish to come back to the problem of ergodicity. We have
seen that the PAEA method is restricted up to now to small pivot
transformations with $N=4,5$. It is thus licit to ask if 
the  danger arises, that with that 
method some relevant subset of polymer conformations is neglected.
With this purpose in mind, we compared the results of the PAEA
calculations of 
the density of states with those obtained using the knot invariant
$\varrho(C)$ in order to distinguish the topology. With this knot
invariant, it is in fact possible to consider pivot transformations
involving an arbitrary number of segments as explained before.
We have found that the densities of states computed with the PAEA method
and with the invariant $\varrho(C)$ are in complete agreement with
each other. In
Fig.~\ref{dost50} it is shown for example the case of the densities of
states for a trefoil knot of length $L=120$ computed with the PAEA
method (grey
squares) and with the 
invariant $\varrho(C)$ evaluated using  the Monte Carlo integration
algorithm of Section~\ref{sectionIII}
 (black circles). As it is possible to realize, both densities of
states coincide. 



\section{Conclusions}\label{sectionVI}
In this work we have computed the specific energy, the specific heat
capacity and  
the radius of gyration of the trefoil knot $3_1$, the figure-eight
$4_1$ 
and the knot $5_1$. Polymer knots have been generated on a simple
cubic lattice
and the conformations  
 used in the Wang-Landau algorithm to get the density of states
$\Omega_m$ have been sampled by means of pivot transformations. The
topology of 
polymer knots needs to be preserved during the sampling procedure. To
this purpose, 
the knot invariant $\varrho(C)$ has been employed. We
have considered  both
attractive and
repulsive 
short-range interactions 
between monomers. 
First, we have analyzed the simplest possible knot $3_1$ at various
lengths $L=50,70,90,120$.  
It has been observed that the specific energy of
polymer knots 
is increasing with the increasing of the temperature irrespective of
the fact that the interactions are 
attractive or repulsive. This is an expected result, because
higher
energy states should be reached
as the temperature increases. 
Less intuitive is that, at high temperatures,
the specific energy of a knot in a given topological configuration
grows with the increasing of the polymer length
in the attractive case as it is illustrated in Fig.~\ref{fig1}(a).
This behavior has been explained in Section~\ref{sectionV} by the
following
two facts: 1) attractive forces become negligible at high temperatures
with respect 
to thermal fluctuation and 2) the effects of being knotted become less
and less important with increasing polymer lengths.
In a similar way, it has been possible to explain the decreasing 
shown in Fig.~\ref{fig1-r}(a)
of the specific energy with the polymer length at high temperatures in
the repulsive case, see Section~\ref{sectionV}. 
In correspondence with the behavior of the specific energy, 
at high temperatures
also the
specific heat capacity 
of longer polymers is larger than that of shorter polymers
in the attractive case, see Fig.~\ref{fig1}(b), while it is smaller in the repulsive case, see Fig.~\ref{fig1-r}(b).
The specific heat capacity
exhibits a very sharp peak in the temperature interval that has been
considered. 
Similar peaks have been observed in the specific heat capacity of
single open linear chains for very short-range attractive interactions
\cite{binder,binder2,kolli}, in knots~\cite{swetnam} and in
star polymers
\cite{wanghe}. An analytical investigation of these phenomena can be
found in \cite{unconventional}.
In the attractive case, see Section~\ref{sectionV}, the peak of the heat capacity is apparently related to the phase
transition of knots from a frozen crystallite state to an expanded
state, similar to that of  a single polymer chain
\cite{binder,binder2}.
In the repulsive case, the interpretion of the peaks is more
complicated and is probably due to a lattice artifact as explained in \cite{yzff}.

The behavior of the mean square radius of gyration $\langle R_G^2
\rangle$ is  displayed in Figs.~\ref{fig3}(a) and \ref{fig3}(b) for
the attractive and repulsive cases respectively. 
It turns out that
longer polymers have bigger mean square gyration radius independently
of the fact that the interactions are repulsive or 
attractive. Indeed, as it is intuitive,
longer polymers should occupy larger volumes.
Our simulations show also that,
in the
attractive case, $\langle R_G^2\rangle$ increases with the increasing
of the
temperature while, on the contrary, it decreases in the repulsive case.
This phenomenon is connected with the analogous increasing and
decreasing of the specific energy mentioned before. As a
matter of fact, shorter radii of gyration imply  
higher specific energies in the repulsive case and lower specific
energies in the attractive case.

The influences of topology on the thermal
properties of polymer knots
have been studied by comparing knots of different types but of
the  same length. In Figs.~\ref{fig2}, \ref{fig2-a} and \ref{fig3b}
are respectively
reported the  data of the specific energy, heat capacity and
gyration radius for the knots $3_1,4_1$ and $5_1$ with length $L=70$.
As it can be observed from Fig.~\ref{fig2}, in
the attractive case
the specific energy and heat capacity decrease  with the knot
complexity. This is due to the fact 
that with the increasing of the complexity of the
 topological configuration while  
the knot length is kept fixed, the conformation of the polymer becomes
more compact 
and thus the number of contacts between the monomers becomes larger.
The opposite situation is observed in the repulsive case. These
results are in agreement with the conclusions of
Refs.~\cite{yzff} and \cite{yf}.  

Beside studying the thermodynamic properties of polymer knots,
the second purpose of this work has been a check of the feasibility
of the use in numerical simulations of knot invariants
which are given in the form of multiple contour integrals.
For this reason, we have considered  one of the simplest knot
invariant, namely the quantity $\varrho(C)$ which is related to the
second coefficient of the Conway polynomial.
The most serious disadvantage of this kind of invariants is that
the evaluation of the multiple integrals is time consuming even within
the Monte Carlo approach adopted here.
However, one should keep in mind that these knot invariants can be
applied to any spatial curve $C$ 
representing the knot under investigation, not necessarily defined on
a cubic lattice or with segments that are straight lines.
This allows to reduce the number of segments composing the polymer
considerably. For example, it is possible to replace in the knot trajectory
up to three segments with a single segments, reducing in this way the
polymer length by a factor three. Up to now, our method  can be efficiently
exploited with polymer knots with lengths $L\le 360$.
Another possibility to reduce the computation time comes from the fact
that one of the integrations in the component $\varrho_2(C)$ of the 
knot invariant  $\varrho(C)$ can be performed analytically.
In this way, the time for calculating $\varrho_2(C)$, which scales with
the polymer length as $t\sim L^4$, may be reduced to $t\sim L^3$.
While there are faster algorithms to preserve the topology, like for
instance the PAEA method discussed in \cite{yzff}, which is both exact
and very fast for small pivot transformations involving up to $N=5$
segments, the use of knot invariants in the form of multiple contour
integrals has the advantage that it works with any number of segments
$N$.
This allows  a faster equilibration of the polymer starting from a
seed configuration and limits the number of sampling during the
calculations with the Wang-Landau Monte Carlo algorithm.

In the future, we plan to extend the present approach, which is valid
for knots, to the case of three linked
polymers.
This is possible because the triple Milnor linking invariant
\cite{milnor}, which is able to distinguish the topology of a link formed
by three knots, is composed by four integrals that have the same
tensor structure of the two components $\varrho_1(C)$ and
$\varrho_2(C)$
of $\varrho(C)$, see for example \cite{Pinedaleal}.

\begin{appendix}  
\section{}

As shown in  \cite{GMMknotinvariant}, in the contribution
$\varrho_1(C)$
to the knot invariant $\varrho(C)$ the integration over the
variable $\vec{\omega}$ in Eq.~(\ref{z2}) can
be performed exactly giving as a result: 
\begin{equation}
\varrho_1(C)=-\frac{1}{32\pi^3}\oint dx^{\mu} 
\int^x dy^{\nu} \int^y dz^{\rho} H_{\mu,\nu,\rho}(\vec{y}-\vec{x},\vec{z}-\vec{x})
\end{equation}
After putting
$a=\vec{y}-\vec{x}$ 
and $b=\vec{z}-\vec{x}$, the tensor $H_{\mu,\nu,\rho}(a,b)$ may be
expressed as follows:
\begin{widetext}
\begin{eqnarray}
\!\!\!\!\!\!&&H_{\mu,\nu,\rho}(a,b)=C_1 C_2 C_3 
\left [\delta_{\nu \rho}(a-b)_{\mu}+\delta_{\mu \rho}
  b_{\nu}-\delta_{\mu \nu} a_{\rho}\right] \nonumber\\ 
\!\!\!\!\!\!&&-C_1 C_2^2 C_3 \epsilon_{\lambda\sigma\tau} a_{\sigma}
b_{\tau} 
\left[
\epsilon_{\rho\mu\alpha} \delta_{\lambda\nu}
\left(
a_{\alpha}+b_{\alpha}\frac{|a|}{|b|}
\right)
+\epsilon_{\nu\mu\alpha}
\delta_{\lambda\rho}
\left(
b_{\alpha}+a_{\alpha}\frac{|b|}{|a|}
\right)
\right]\nonumber\\
\!\!\!\!\!\!&&+C_1 C_2 \epsilon_{\lambda\sigma\tau} a_{\sigma} b_{\tau}
 \left\{
\epsilon_{\rho\mu\alpha} \delta_{\lambda\nu} 
\left[
b_{\alpha} 
\frac{|a-b|-|a|}{b^2}+(a-b)_{\alpha} \frac{|a|+|b|}{(a-b)^2}
\right]
+\epsilon_{\nu\mu\alpha} \delta_{\lambda\rho} 
\left[
a_{\alpha} \frac{|a-b|-|b|}{a^2}+(b-a)_{\alpha}
  \frac{|a|+|b|}{(a-b)^2}
\right] 
\right\}\nonumber\\
&&
\end{eqnarray}
\end{widetext}
where
\begin{eqnarray}
C_1(a,b)=\frac{2\pi}{|a||b||a-b|}\\
C_2(a,b)=\frac{1}{|a||b|+a_{\mu}b_{\mu}}\\
C_3(a,b)=|a|+|b|-|a-b|
\end{eqnarray}
$|a|,|b|$ and $|a-b|$ mean the module of the variable $a,b$ and $(a-b)$, like $|a|=\sqrt{(y-x)^2_1+(y-x)^2_2+(y-x)^2_3}$. $a_{\alpha},b_{\alpha},(a-b)_{\alpha}$ mean the $\alpha$-th component of the corresponding variable, $\alpha=1,2,3$.
\end{appendix}
\begin{acknowledgments} 
The support of the Polish National Center of Science,
scientific project No. N~N202~326240, is gratefully acknowledged.
The simulations reported in this work were performed in part using the HPC
cluster HAL9000 of the Computing Centre of the Faculty of Mathematics
and Physics at the University of Szczecin.
\end{acknowledgments}

\end{document}